%% file: main.tex
\documentclass{article}
\usepackage{stylefile}

\title{Inference in the presence of model-form uncertainties: Leveraging a prediction-oriented approach to improve uncertainty characterization}
\author{Rebekah White, Rileigh Bandy, \& Teresa Portone}
\date{January 2025}

\begin{document}

\maketitle

\input{abstract}
\input{intro}

\input{background}

\input{methods}
\input{computational_results}
\input{conclusions}
\input{acknowledgements}

\pagebreak

\bibliographystyle{plain}
\bibliography{MFU_inf.bib}

\end{document}

%% file: abstract.tex
\begin{abstract}
    Bayesian inference is a popular approach to calibrating uncertainties, but it can underpredict such uncertainties when model misspecification is present, impacting its reliability to inform decision making.
    Recently, the statistics and machine learning communities have developed prediction-oriented inference approaches that provide better calibrated uncertainties and adapt to the level of misspecification present.
    However, these approaches have yet to be demonstrated in the context of complex scientific applications where phenomena of interest are governed by physics-based models. 
    Such settings often involve single realizations of high-dimensional spatio-temporal data and nonlinear, computationally expensive parameter-to-observable maps.
    This work investigates variational prediction-oriented inference in problems exhibiting these relevant features; namely, we consider a polynomial model and a contaminant transport problem governed by advection–diffusion equations. 
    The prediction-oriented loss is formulated as the log-predictive probability of the calibration data. 
    We study the effects of increasing misspecification and noise, and we assess approximations of the predictive density using Monte Carlo sampling and component-wise kernel density estimation.
    A novel aspect of this work is in applying prediction-oriented inference to the calibration of model-form uncertainty (MFU) representations, which are embedded physics-based modifications to the governing equations that aim to reduce (but rarely eliminate) model misspecification.
    The computational results demonstrate that prediction-oriented frameworks can provide better uncertainty characterizations in comparison to standard inference while also being amenable to the calibration of MFU representations.
\end{abstract}

%% file: intro.tex
\section{Introduction}\label{sec:intro}

Computational modeling and simulation are increasingly used to inform decision making in scenarios where observational data may not be available. 
However, confidence in such decisions must be established through uncertainty quantification in calibration scenarios where observational data are available.
While Bayesian inference provides a natural framework for characterizing uncertainties from data, it is well known to underpredict uncertainties when model misspecification is present due to undesirable posterior contraction~\cite{bjk_kleijn_bernstein-von-mises_2012}.
Such behavior poses significant challenges in the context of physics-based modeling of complex phenomena, wherein simplifying assumptions are often necessary due to limited knowledge or computing power. 
Thus, some degree of model misspecification is often present, making inference approaches amenable and adaptable to such misspecification essential to leveraging modeling and simulation to inform decision making.

Alternative probabilistic frameworks have been proposed to improve the characterization of uncertainties when model misspecification is present. 
One class of approaches can be broadly categorized as generalized Bayesian inference (GBI) whose solution is commonly referred to as the Gibbs posterior~\cite{bissiri_general_2016,Zhang2006b,zhang_e-entropy_2007,jiang_gibbs_2008}.
Here, the notion of adherence of model predictions to data is relaxed from the expected log-likelihood found in traditional Bayes, instead considering generalized expected losses, whose flexibility can provide robustness to misspecification and be tailored to the problem at hand.
GBI encompasses and relates to a variety of approaches including tempered posteriors~\cite{Grunwald2014}, Safe Bayes~\cite{Grunwald_2012}, coarsened posteriors~\cite{miller_robust_2019}, and approximate Bayesian computation (ABC)~\cite{marin_approximate_2012,sisson2018}. 
Additionally, while borne out of differing motivations in the machine learning community, generalized solutions can be viewed as minimizers to probably approximately correct (PAC) Bayes upper bounds on the generalization errors of stochastic predictors~\cite{mcallester_pac-bayesian_1999,alquier_pac-bayesian_2008,mcallester_pac-bayesian_2003,mcallester_1999_model_av,guedj2019primerpacbayesianlearning}.
Although GBI can improve uncertainty characterizations when misspecification is present for finite data, they do not necessarily prevent unwanted contraction of the posterior and, as result, can still provide posteriors with poor predictive densities~\cite{mclatchie_predictively_2025}.
Furthermore, in many settings, such as ABC, robustness of the approach may rely on proper specification of robust summarizing statistics, which may not be feasible in practice.

Therefore, this work considers alternative inference approaches that are robust to misspecification, whose solutions (posteriors) provide predictive uncertainties that assign high probability to the calibration data, while not being so broad as to be uninformative. 
To do so, we leverage a prediction-oriented framework recently developed in the statistics and machine learning communities~\cite{mclatchie_predictively_2025,morningstar_2022,shen_prediction-centric_2025}.
Similar to GBI~\cite{knoblauch2022_GVI}, the inference problem is cast as an infinite-dimensional optimization problem admitting variational approximations, where the objective is comprised of a loss term measuring adherence to the data as well as a regularization term measuring adherence to prior beliefs. 
Here, however, the loss is formulated to directly target how well predictive uncertainties characterize the observational data by considering a scoring rule that acts directly on the predictive density. 
While conceptually appealing, there have been limited explorations into applying these prediction-oriented inference techniques to complex science applications where phenomena of interest are governed by potentially large-scale physics-based models.
Thus, the aim of this work is to investigate performance of the approach in contexts relevant to such problems.
Specifically, we are interested in scenarios where 1) observations consist of high-dimensional,  spatio-temporal data; 2) parameter-to-observable maps are nonlinear and defined by computationally expensive, differential equation–based models; and 3) likelihoods are available only implicitly. 
Moreover, a feature unique to physics-based modeling is the ability to leverage representations of model-form uncertainty (MFU)~\cite{portone_quantifying_2025,bandy_quantifying_2024,oliver_validating_2015,morrison_representing_2018}, which are embedded physics-based modifications to the governing equations, to help mitigate misspecification.
Novel in this work, we explore the application of prediction-oriented inference for informing such uncertainty representations alongside model parameters.

In applying prediction-oriented inference to computational examples demonstrating features relevant to science and engineering applications, we leverage the log-predictive probability of the data as the loss function in the variational objective.
We consider a polynomial example that admits closed-form representations of the log-predictive loss as well as a contaminant transport problem, governed by advection-diffusion partial differential equations (PDEs).
With these computational examples, we explore the impact of increasing misspecification and noise on the predictive inverse problem solutions in comparison to standard variational inference. 
Additionally, we numerically assess convergence of Monte Carlo (MC) estimators of the predictive density in comparison to the expected log-likelihood employed in standard approaches. 
We further investigate the impact of component-wise density approximations often required for high-dimensional data where explicit likelihoods are unavailable. 
For the contaminant transport problem, we demonstrate how informing MFU representations via the prediction-oriented framework can provide better predictive uncertainties and result in less biased parameter estimates in comparison to standard approaches.
Overall, this work serves as a preliminary investigation into the application of prediction-oriented inference to physics-based models of complex phenomena and the challenges therein. 
The primary contributions of this work are: 
\begin{itemize}
    \item Application of prediction-oriented inference to problems with features relevant to real-world engineering and science problems. 
    \item Exploration into the impacts of various approximations often required for high-dimensional data and complex physics-based models.
    \item Novel application of prediction-oriented frameworks to the calibration of MFU representations alongside model parameters.
\end{itemize}

The remainder of this paper is organized as follows. 
First,~\Cref{sec:background} reviews standard Bayesian inference and resulting implications in the presence of misspecification in~\Cref{sec:bayes} followed by generalized Bayesian inference and related approaches in~\Cref{sec:gen_bayes}.
In~\Cref{sec:methods}, we review the methods employed in the computational demonstrations, with a discussion of prediction-oriented inference and relevant approximations given in~\Cref{sec:pip} and an overview of MFU representations provided in~\Cref{sec:mfu}.
The computational results are presented in~\Cref{sec:comp_res}, where~\Cref{sec:poly} considers a polynomial model, while~\Cref{sec:ade} leverages a PDE-based model of contaminant transport through a heterogeneous porous medium and explores the impact of leveraging MFU representations in the prediction-oriented inverse problem.
Final conclusions are provided in~\Cref{sec:conclusions}.

%% file: background.tex
\section{Background on standard and generalized Bayesian inference}\label{sec:background}

We first introduce the standard Bayesian inference paradigm and discuss the associated challenges when performing inference in the presence of model misspecification in~\Cref{sec:bayes}. 
Then, in~\Cref{sec:gen_bayes}, we discuss generalized Bayesian inference and related approaches that have been proposed to better enable calibration in the context of model misspecification. 
While we ultimately find that such generalized approaches do not provide the panacea for calibrating uncertainties when misspecification is present, we discuss such approaches as they provide important context for the more recently-developed prediction-oriented frameworks discussed in~\Cref{sec:pip}.


\subsection{Standard Bayesian inference and implications in the presence of model misspecification}\label{sec:bayes}

The standard Bayesian inference paradigm is a popular approach for informing model uncertainties as it provides not only estimates for the parameters $\param = [\lcparam_1, \dots, \lcparam_{\dimparam}]^{\top}$ but corresponding uncertainty characterizations that can inform decision making.
First, consider the parameter-to-observable map $f({\bm x}; \param):\Param \to \mathcal{Y}$ given in a general form where ${\bm x}$ represents a vector of discretized spatial locations, and $\Param$ is assumed to be finite dimensional.
Often, the parameter-to-observable map is constructed using an observation operator acting on the solution of the underlying model (e.g. differential equations).
The goal of inference is to leverage observational data $\data = [y_1, \dots, y_{n_y}]^{\top} \in \mathcal{Y} \subset \mathbb{R}^{n_y}$ to inform the model parameters $\param$.
When formulating the Bayesian inverse problem, one specifies a statistical model, which represents the data collection procedure. 
Often the following additive Gaussian noise assumption is leveraged:
\begin{eqnarray}\label{eq:stat_mod}
    \data({\bm x}) = f({\bm x}; \param) + {\bm \epsilon}, \quad {\bm \epsilon} \sim \mathcal{N}\left({\bm 0}, {\bm \Gamma}_{\text{noise}}\right),
\end{eqnarray}
where ${\bm \epsilon}$ represents noise realizations, whose distribution is defined by the noise covariance ${\bm \Gamma}_{\text{noise}}$.
The form of~\eqref{eq:stat_mod} leads to the likelihood specification given by
\begin{eqnarray}\label{eq:like}
    \likedens(\data | \param) \sim \mathcal{N}\left( f({\bm x}; \param), {\bm \Gamma}_{\text{noise}}\right).
\end{eqnarray}
The solution to the Bayesian inverse problem is a posterior probability density function on the model parameters computed according to  Bayes' rule as
\begin{equation}\label{eq:bayes_stand}
    \postdens(\param | \data) \propto \likedens(\data | \param) \priordens(\param),
\end{equation}
where $\priordens(\param)$ is the prior density, reflecting one's beliefs regarding parameter uncertainties prior to collecting any data.
To visualize how uncertainty in the model parameters propagates to uncertainty in a model-prediction quantity of interest (QoI), one can leverage the posterior pushforward (pf) distribution.
Defining this formally, let the QoI be defined as $g(\param):\Param \to \mathcal{G}$, and define $G:=g(\param)$. 
Uncertainty in $G$ is induced by $\postdens(\param | \data)$ and is characterized by the pushforward measure, whose density we assume exists and is continuous with the Lebesgue measure on $\mathcal{G}$, which is defined as 
\begin{eqnarray}\label{eq:post_pushforward}
    \pi_{\text{pf}}(z | \data) = \int_{\Param} \delta(
    z - g(\param))\postdens(\param | \data) d\param.
\end{eqnarray}
To visualize this pushforward density, one can consider the marginal $100(1-\alpha)\%$ credible intervals for component $G_j$ defined as 
\begin{eqnarray}\label{eq:cred_int}
    C_{\alpha}^{(j)} = \left[ z_{\frac{\alpha}{2}}^{(j)}, z_{1-\frac{\alpha}{2}}^{(j)}\right],
\end{eqnarray}
where $z_{\frac{\alpha}{2}}^{(j)}$ and $z_{1-\frac{\alpha}{2}}^{(j)}$ respectively denote the $\frac{\alpha}{2}$ and $1-\frac{\alpha}{2}$ quantiles of the marginal predictive distribution.

%

While it is true that incorrect models can still be useful for understanding phenomena of interest, practical issues often arise when leveraging such imperfect models in inference. 
Implicit in the likelihood specification \eqref{eq:like} is the assumption that the mathematical model is capable of representing the data-generating distribution. 
However, when model misspecification is present, such an assumption no longer holds, which can significantly impact the resulting uncertainty characterization.
Following~\cite{scarinci_bayesian_2021}, we define model misspecification generally and discuss its implications in the context of inference.
Consider observations $y_{1:n}$ drawn independently and identically from an unknown data-generating distribution with joint density $\pi_g(y_{1:n})$.
In practice, inference is conducted using a parametric family $\{\pi_p(y_{1:n} | \lcparam): \lcparam \in \Param \}$ intended to approximate this distribution.
The model is said to be well-specified if the true distribution lies within this family, i.e. if there exists $\lcparam_0 \in \Param$ such that $\pi_g(y_{1:n}) = \pi_p(y_{1:n} | \lcparam_0)$; otherwise, if $\pi_g(y_{1:n}) \neq \pi_p(y_{1:n} | \lcparam_0)$ for all possible values of $\lcparam_0$, the model is misspecified.
Under standard regularity conditions, the Bernstein-von Mises theorem implies that for well-specified models, the posterior distribution converges to a Gaussian centered at the data-generating parameter $\lcparam_0$ as the sample size grows.
In contrast, when misspecification is present, the posterior instead concentrates around the minimizer of the Kullback-Leibler (KL) divergence given as 
\begin{eqnarray}\label{eq:kl_min}
    \lcparam^* = \operatorname*{argmin}_{\lcparam \in \Param} \mathcal{D}_{KL} \left(\pi_g(\cdot) \,||\, \pi_p(\cdot | \lcparam)\right).
\end{eqnarray}
See~\cite{bjk_kleijn_bernstein-von-mises_2012} for more details.
The minimizer given in~\eqref{eq:kl_min} need not be unique and does not, in general, correspond to a model capable of reproducing the observed data.
Nevertheless, the posterior concentrates about this value with increasing amounts of data, leading to increased confidence about an incorrect model.
In practice, this phenomenon results in the propagated (often extrapolated) posterior uncertainties being unreliable to inform decision making. 

\begin{figure}[h]
    \centering
    \includegraphics[scale=0.48]{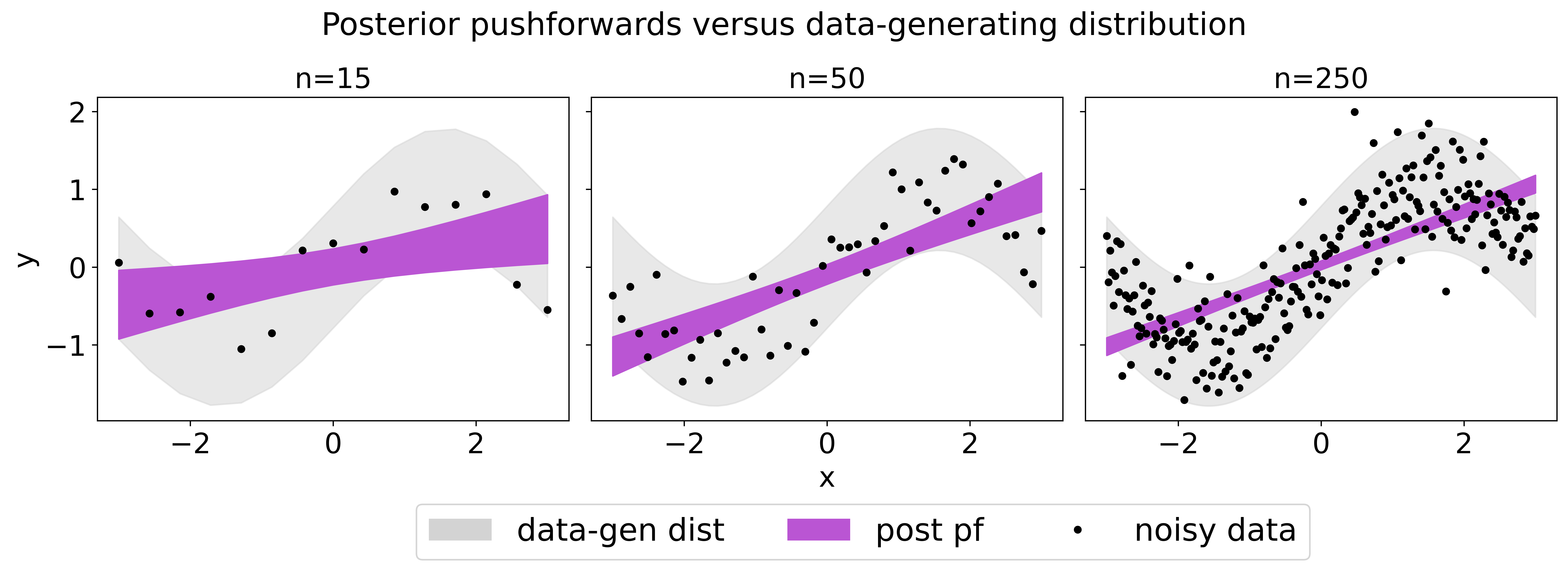}
    \caption{Comparison of the $95\%$ marginal credible intervals corresponding to the true data-generating (sinusoidal) distribution versus posterior pushforward for varying amounts of data $y_{1:n}$, for $n=15, 50, 250$ (left to right), where the inference model (linear) is misspecified. 
    }
    \label{fig:model_misspecification}
\end{figure}

To illustrate model misspecification and its impact on the resulting standard Bayesian posterior, consider that the noisy data are determined according to the following true data-generating process   
\begin{eqnarray}
    y^{\text{obs}} = \sin(x) + \epsilon, \quad \epsilon \sim \mathcal{N}(0, 0.4^2).
\end{eqnarray}
However, due to misspecification in the inference model, the assumed statistical model is given by 
\begin{eqnarray}
    y = ax + b + \epsilon, \quad \epsilon \sim \mathcal{N}(0, 0.4^2).
\end{eqnarray}
\Cref{fig:model_misspecification} depicts the $95\%$ credible intervals of the true data-generating distribution versus the posterior pushforward, where the posterior is computed with increasing amounts of data $n=15, 50, 250$.
From~\Cref{fig:model_misspecification}, one can see that as the sample size grows, the posterior and hence its pushforward becomes more concentrated about the misspecified model, making the uncertainty characterization unreliable.
While in practice, one would not have access to the true data-generating distribution for direct comparison, misspecification can be indicated by poor coverage of propagated uncertainties and asymmetric residuals.

\subsection{Generalized Bayesian inference \& related approaches}\label{sec:gen_bayes}

Here, we provide background on \textbf{generalized Bayesian inference (GBI)} and related approaches aimed at increasing robustness of the inference problem to model misspecification.
First, consider that the standard Bayesian posterior can be viewed as the solution to the following infinite-dimensional optimization problem 
\begin{eqnarray}\label{eq:bayes_opt}
    \postdens(\param | \data) = \operatorname*{argmin}_{q\in \mathcal{P}(\Param)} 
    \mathbb{E}_{q(\param)} \left[ -\log(\likedens(\data | \param))\right] + \text{KL}\left( q(\param) || \priordens(\param) \right),
\end{eqnarray}
where $\mathcal{P}(\Param)$ is the space of all probability measures on $\Param$~\cite{bissiri_general_2016, knoblauch2022_GVI}. 
The first element of the objective function represents fidelity to data by measuring the loss given by the log-likelihood; the second term represents fidelity to prior beliefs.
From this optimization-centric view of inference, alternative loss functions $l(\param, \data)$ can be used in place of the log-likelihood to provide robustness to model misspecification as follows:
\begin{eqnarray}\label{eq:GBI_opt}
    \postdens(\param | \data) = \operatorname*{argmin}_{q\in \mathcal{P}(\Param)} 
    \mathbb{E}_{q(\param)} \left[ -l(\param, \data) \right] + \text{KL}\left( q(\param) || \priordens(\param) \right),
\end{eqnarray}
The work of~\cite{bissiri_general_2016}, shows that the solution to the more generalized inference problem~\eqref{eq:GBI_opt} is given as 
\begin{eqnarray}\label{eq:GBI_post}
    \postdens(\param | \data) \propto \exp\left(-l(\param, \data)\right)\priordens(\param),
\end{eqnarray}
which is also known as the Gibbs posterior~\cite{Zhang2006b}, whose robustness to misspecification has been explored~\cite{jiang_gibbs_2008}.
Note that we refer to all solutions to the optimization problems such as~\eqref{eq:bayes_opt} and~\eqref{eq:GBI_opt} as posteriors $\postdens(\param | \data)$ regardless of whether they are the solution to standard or generalized optimization objectives or approximated, for example, through variational approaches.
The work of~\cite{bissiri_general_2016} goes on to show solutions in the form of Gibbs posteriors provide ``coherent, decision theoretic representation of posterior beliefs under model misspecification'' while also providing a way to scale the information in the data to the prior information by leveraging a scaling parameter $\alpha$ when computing the posterior as follows:
\begin{eqnarray}\label{eq:scaled_gibbs}
    \postdens(\param | \data) \propto \exp\left(-\alpha \cdot l(\param, \data)\right) \priordens(\param), \quad \alpha > 0.
\end{eqnarray}

GBI encompasses or relates to a variety of alternative approaches. 
Consider, first, that if one defines the loss in~\eqref{eq:GBI_post} as the log-likelihood, the standard Bayesian posterior is recovered.
Leveraging the log-likelihood in the scaled Gibbs posterior~\eqref{eq:scaled_gibbs} results in a \textbf{tempered posterior}, which can reduce the influence of the misspecified model by
downweighting the likelihood; see~\cite{Grunwald2014} for more details on the robustness of tempered posteriors to model misspecification.
\textbf{Safe Bayes} also leverages~\eqref{eq:scaled_gibbs} but adaptively estimates the temperature parameter $\alpha$ during inference by considering predictive performance as measured by the posterior expected loss~\cite{Grunwald_2012}.
The effect of this adaptive optimization of the tempering parameter is that when misspecification is high, $\alpha$ decreases to reduce the influence of the likelihood.
The Gibbs posterior also relates to an alternative approach for increasing robustness of the Bayesian posterior, termed \textbf{coarsening}~\cite{miller_robust_2019}.
Rather than conditioning on the event that the model generated the data, coarsened posteriors condition on the event that the empirical distribution on the observational data is close, in some statistical sense, to the empirical distribution of the data generated by the model. 
The Gibbs posterior provides a large-sample approximation to a coarsened posterior.

There also exist connections between GBI and techniques developed in the context of machine learning (ML), namely probably approximately correct (PAC) learning~\cite{valient_1984}, where the aim, roughly speaking, is to provide theoretical guarantees that an ML model (trained on finite data) will generalize well to new, unseen data.
The Bayesian extensions of this framework, \textbf{PAC-Bayes}~\cite{mcallester_pac-bayesian_1999,alquier_pac-bayesian_2008,mcallester_pac-bayesian_2003,mcallester_1999_model_av,guedj2019primerpacbayesianlearning}, considers generalization guarantees for stochastic predictors without assuming a particular model for the data generating process~\cite{alquier_properties_2016}.
Such approaches form upper bounds on the empirical risk (generalization error); it can be shown that the Gibbs posterior is the minimizer to the upper bound~\cite{mcallester_pac-bayesian_2003, alquier_properties_2016}. 
Although the motivation behind PAC-Bayes is not necessarily concerned with the properties of the resulting posterior distribution, the connections to GBI indicate that explorations into computationally tractable PAC upper bounds could improve computation of GBI solutions.

Although motivated by intractable rather than unreliable likelihoods, \textbf{approximate Bayesian computation (ABC)}~\cite{marin_approximate_2012,sisson2018} is conceptually similar to GBI.
By leveraging flexible notions of how close corresponding model evaluations (or summarizing statistics) are to the observational data, ABC provides samples from an approximate posterior without likelihood evaluation.
The work of~\cite{schmon2021} explores how GBI can be used to design ABC algorithms robust to model misspecification by considering distance metrics based on Wasserstein distance and soft-thresholding. 
Furthermore, the choice of summarizing statistic can also provide robustness to misspecification~\cite{Lewis2021}, although determining such statistics is highly application specific.
%
Lastly, we note recent work that further generalizes~\eqref{eq:GBI_opt} by considering statistical divergences beyond just the KL-divergence\footnote{The loss formulations of~\cite{knoblauch2022_GVI} are written as a summed loss as they consider multiple independent and identically distributed realizations of data, whereas $\data$ in~\eqref{eq:GBI_opt} is one multivariate realization of data.}~\cite{knoblauch2022_GVI}.
This approach is termed \textbf{generalized variational inference (GVI)} as $\mathcal{P}(\Param)$ is restricted to a variational family of probability measures as such generalizations may not provide closed-form representations of the posterior as in~\eqref{eq:GBI_opt} and~\eqref{eq:scaled_gibbs}.
Advocating for an optimization-centric view of Bayesian inference, GVI aims to address model and prior misspecification, while maintaining computational tractability through variational approximations.

Although the origins and motivations behind the aforementioned approaches may differ, they can all be shown to provide some level of robustness to model misspecification in comparison to standard Bayesian paradigms. 
Such approaches do not, however, directly penalize predictive uncertainty characterizations and hence do not ensure resulting posteriors are capable of describing the calibration data in the sense of assigning it high probability.
In fact, as noted in~\cite{mclatchie_predictively_2025}, both standard and generalized Bayesian inference result in posterior contraction with increased amounts of data and can therefore lead to poor predictives.
As such, we instead consider prediction-oriented inference~\cite{mclatchie_predictively_2025}, wherein the loss terms are formulated to specifically target predictive how well predictive uncertainties describe the observational data. 
We discuss this approach in detail in the following section.

%% file: methods.tex
\section{Methods}\label{sec:methods}

Here, we describe the methods employed in the computational results that follow in~\Cref{sec:comp_res}. 
First, we discuss the prediction-oriented framework following the works of~\cite{mclatchie_predictively_2025, morningstar_2022,shen_prediction-centric_2025,lai2025predictivevariationalinferencelearn}.
Then, in~\Cref{sec:mfu}, we present the relevant background on MFU representations highlighting existing approaches to calibrating such representations and the challenges therein.

\subsection{Prediction-oriented inference}\label{sec:pip}
Following~\cite{mclatchie_predictively_2025,morningstar_2022}, we provide the approach for prediction-oriented inference that is leveraged in the computational results following in~\Cref{sec:comp_res}.
The general form of the prediction-oriented optimization objective is
\begin{eqnarray}\label{eq:PIP}
    \postdens(\param | \data) = \operatorname*{argmin}_{q\in \mathcal{Q}(\Param)} 
    \mathcal{L}(\param, \data) + \text{KL}\left( q(\param) || \priordens(\param) \right),
\end{eqnarray}
where $\mathcal{L}$ is a loss quantifying how well the  predictive distribution describes the observational (calibration) data and $\mathcal{Q}(\Param) \subset \mathcal{P}(\Param)$ is a variational family of probability densities. 
The general form for the predictive distribution on a QoI is defined similarly to the pushforward in~\Cref{sec:bayes}, where again $g(\param):\Param \to \mathcal{G}$, but the noise model results in the random variable associated with observations being given as $G | \param \sim \pi(\cdot | \param)$. 
The corresponding posterior predictive is then defined as 
\begin{eqnarray}\label{eq:post_predictive}
    \pi_{\text{pred}}(z) = \int_{\Param} \pi(z | \param)q(\param) d\param, \quad \text{for}.
\end{eqnarray}
To visualize the predictive distribution, one can leverage the marginal credible intervals defined in~\eqref{eq:cred_int}.
In prediction-oriented inference, one is interested in how well the posterior predictive describes the observational data $\data$. 
Therefore, following~\cite{mclatchie_predictively_2025}, we formulate the prediction-oriented loss based on the probability of the data under the posterior predictive given by
\begin{eqnarray}\label{eq:poly_gen_loss}
    \mathcal{L}(\param, \data) = - \log \pi_{\text{pred}}({\bm y}) = - \log \int_{\Param} \likedens( {\bm y} | \param) q(\param) d\param = - \log \mathbb{E}_{q(\param)} \left[ \likedens( {\bm y} | \param)\right].
\end{eqnarray}
The logarithmic scoring rule provides the measure of accuracy of the predictive distribution.
Note that prediction-oriented posteriors can be computed with alternative scoring rules such as maximum mean discrepancy or cumulative ranked probability score; see~\cite{mclatchie_predictively_2025,shen_prediction-centric_2025} for more details.
We refer to solving~\eqref{eq:PIP} as performing \textbf{predictive variational inference (PVI)} and the resulting solution as the \textbf{prediction-oriented posteriors}\footnote{Note that~\cite{mclatchie_predictively_2025} refers to the resulting posteriors as the predictively-oriented (PrO) posteriors.}.

\begin{figure}[h]
    \centering
    \includegraphics[width=0.7\linewidth]{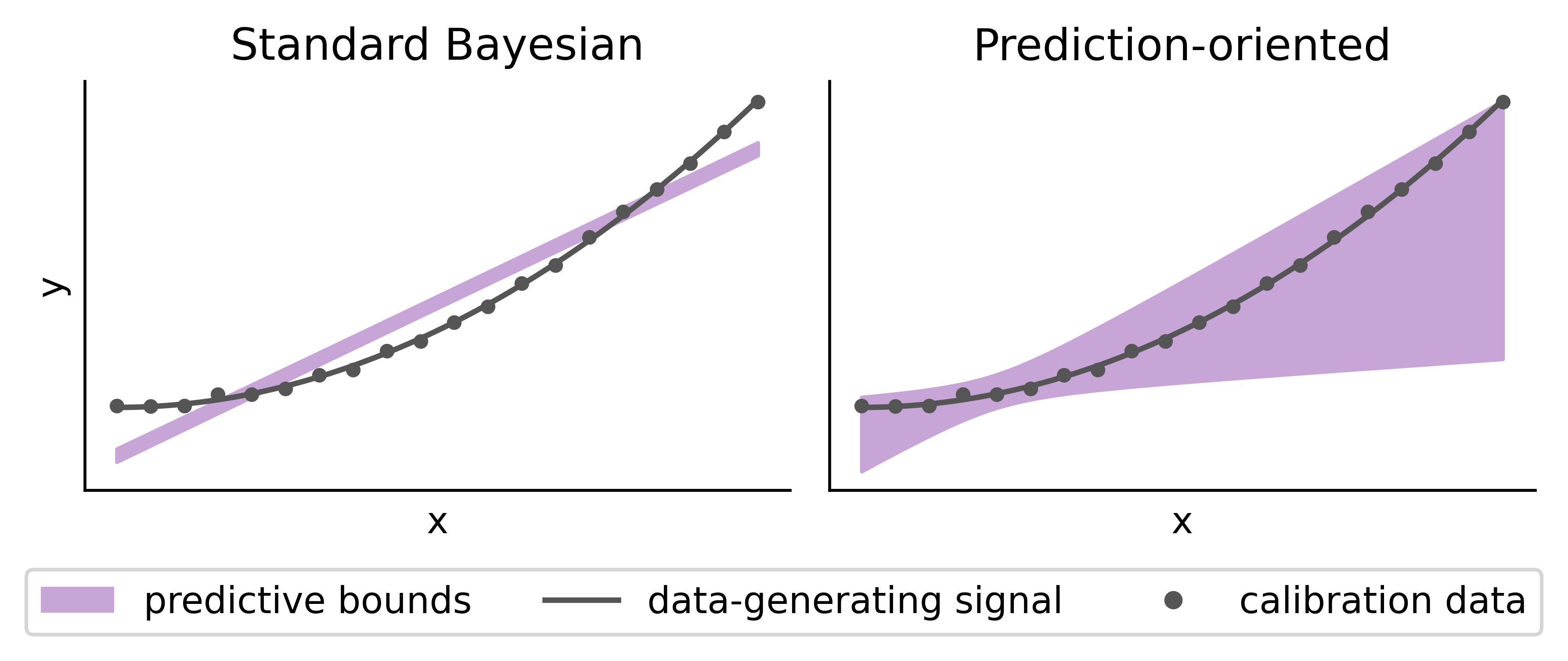}
    \caption{Illustrative depiction comparing standard Bayesian variational inference to predictive variational inference, where the true data-generating signal is quadratic $y = 3x^2 + 1$, while the inference model is misspecified as linear $y = \theta_0 + \theta_1 x$.
    }
    \label{fig:example_notional}
\end{figure}

We note that in comparison to the standard formulation in~\eqref{eq:bayes_opt}, the formulation of~\eqref{eq:poly_gen_loss} materially results in switching the order of the logarithm and expectation. 
Although a seemingly small detail, the difference amounts to whether the focus is inferential or predictive. 
In standard inference, the aim is to determine the best parameters. 
Here, the log-likelihood is a score measuring how well a parameter $\param \sim q(\param)$ describes the data $\data$, and one determines the best solution by computing the \textit{average score of each model induced by a given parameter value}.
In comparison, in predictively-oriented inference, the primary quantity of interest is the posterior predictive distribution $\pi_{\text{pred}}$. 
The score is the log-probability of the predictive, which takes into account the entire distribution $q(\param)$; hence, the best solution is the \textit{score of the averaged model}.
One can also understand the implications of the ordering of the expectation and logarithm by noting from Jensen's inequality that 
\begin{eqnarray*}
- \log \int_{\Param} \likedens( {\bm y} | \param) q(\param) d\param 
\leq 
- \int_{\Param} \log \likedens( {\bm y} | \param) q(\param) d\param,
\end{eqnarray*}
where the bound is tight only if the model is well specified, indicating that
asymptotically, with no misspecification, the solutions to~\eqref{eq:bayes_opt} and~\eqref{eq:PIP} correspond~\cite{masegosa2020learning,morningstar_2022,mclatchie_predictively_2025}.
To provide further intuition,~\Cref{fig:example_notional} presents an illustrative comparison between VI and PVI, where the data-generating signal is quadratic, while the inverse model is misspecified as linear.
\Cref{fig:example_notional} depicts the underprediction in uncertainty associated with standard Bayesian inference, where the calibration (observational) data are assigned low probability by the posterior predictive.
In comparison, the prediction-oriented approach assigns the data high probability, resulting in an uncertainty characterization broad enough to encompass the data, but not too broad as to be uninformative.

This work investigates the suitability of prediction-oriented inference for large-scale engineering applications.
Here, models are typically expensive to compute, commonly governed by PDEs.
Additionally, data are typically limited, where often we have access to only one realization of a high-dimensional dataset.
Such a focus differs significantly from the statistical models and machine learning applications considered in~\cite{mclatchie_predictively_2025}.
Furthermore, our aim is to leverage prediction-oriented posteriors to inform model-form representations alongside model parameters.
As such, it is necessary to consider the practical implications of solving~\eqref{eq:PIP} in such contexts.
Consider first, that even with access to an explicit likelihood function, the loss defined in~\eqref{eq:poly_gen_loss} must always be approximated.
In comparison to standard VI, MC approximation of the log-likelihood is significantly more numerically stable than directly estimating the following marginal likelihood:
\begin{eqnarray}\label{eq:pred_MC}
    \pi_{\text{pred}} \approx \frac{1}{n} \sum_{i=1}^{n}\likedens(\data | \param^{(i)}), \quad \param^{(i)} \sim q(\param),
\end{eqnarray}
which can be numerically more challenging than the marginal log-likelihood found in standard approaches.
Since this approximation occurs at every iteration of the optimization routine, a significant number of model evaluations may be required.
%

In addition to potentially requiring a significant forward model evaluations, the approach given in~\eqref{eq:pred_MC} also depends upon an explicit likelihood function, which is not always available in practice. 
In such cases, one can consider a likelihood-free or implicit likelihood-based approach. 
Rather than directly evaluating the likelihood, one instead samples from the predictive, and leverages density estimation to evaluate the loss given in~\eqref{eq:poly_gen_loss}; see~\Cref{alg:kde_pred} for a detailed description of this approach for additive noise and kernel density estimation (KDE).
\enlargethispage{\baselineskip}
\begin{algorithm}[h]
    \DontPrintSemicolon
    \SetAlgoLined
    \KwInput{Observational data $\data^{\text{obs}} \in \mathbb{R}^{\datadim}$, parameter-to-observable map $f(\param):\Param \to \mathbb{R}^{\datadim}$, noise distribution $\pi_{\epsilon}({\bm \epsilon})$, sampling density $q(\param)$, kernel function $K$, and number of desired samples $n > 0$}
    \textbf{Sample from the predictive:} \vspace{0.3em}\\
    \For{$i=1, \dots, n$}
    {\vspace{0.3em}
        Generate $\param^{(i)} \sim q(\param)$ \\\vspace{0.3em}
        Sample noise ${\bm \epsilon}^{(i)} \sim \pi_{\epsilon}({\bm \epsilon})$ \\\vspace{0.3em}
        Simulate data as $\data^{(i)} = f(\param^{(i)}) + {\bm \epsilon}^{(i)}$
        \\\vspace{0.3em}
        }
    \textbf{Kernel density estimation:} \vspace{0.3em}\\
    Estimate bandwidth $h$\footnotemark
    \\
    $\hat{\pi}_{\text{pred}}({\bm z}) = \frac{1}{nh^{\datadim}} \sum_{i=1}^{n} K\left(\frac{{\bm z} - \data^{(i)}}{h}\right)$
     \\\vspace{0.3em}
    \KwOutput{Loss evaluations $\mathcal{L}(\param, \data^{\text{obs}}) = -\log \hat{\pi}_{\text{pred}}(\data^{\text{obs}})$} 
    \vspace{0.3em}
    \caption{Likelihood-free evaluation of the loss in~\eqref{eq:poly_gen_loss}}
     \label{alg:kde_pred}
\end{algorithm}
\interfootnotelinepenalty=10000
\footnotetext{While one could directly specify a bandwidth, often these are estimated as part of the algorithm. This work leverages the \texttt{scipy.stats.gaussian\_kde} default of Scott's rule~\cite{scipy-gaussian-kde-doc,scott_multivariate_1992}.}
One challenge, however, in applying this approach is that the data is often high-dimensional, for which density estimation scales poorly.  
To enable a tractable approach, we thus approximate the density by leveraging the following component-wise approximation 
\begin{eqnarray}\label{eq:comp_kde}
    K\left(\frac{{\bm z} - \data^{(i)}}{h}\right) = \prod_{j = 1}^{\datadim} K_j \left(\frac{z_j - y_j^{(i)}}{h_j} \right),
\end{eqnarray}
where the subscript $j = 1, \dots, \datadim$ denotes the component of interest; ${\bm z}$ or $z$ denote the argument of the function; and $K$ and $K_j$ respectively denote the multivariate and univariate kernels considered, which are Gaussian  
in the computational results that follow.
While the component-wise approximation will still ensure the predictive describes the data in a marginal sense, the impact on the resulting posterior  could impact the validity of extrapolating the uncertainties. 
Future work could explore alternative approaches to density estimation that scale to moderate dimensions, the benefits of which include not requiring explicit likelihood functions, making them amenable to a broader class of problems.
Furthermore, density estimation provides the full predictive distribution, which once constructed, can efficiently be evaluated for varying data collection strategies at negligible additional computational cost; this could be incredibly useful in optimal experimental design, for example.

\subsection{Model-form uncertainty representations}\label{sec:mfu}

A novel aim of this work is to evaluate the performance of PVI in the context of calibrating model-form uncertainty (MFU) representations.
Model misspecification often arises from simplifying model assumptions (e.g.~closure/constitutive models, neglected interactions, assumed isotropy), which result from limited knowledge of the underlying process or limited ability to model such complexities in a computationally tractable way.
MFU representations aim to better represent the uncertainties associated with such assumptions to improve predictive and extrapolative performance of the models~\cite{oliver_validating_2015,morrison_representing_2018,portone_representing_2019,bandy_quantifying_2024,bandy_stochastic_2025,portone_quantifying_2025}.
Such representations are formulated as parameterized modifications to modeling assumptions. They can be described in a general mathematical form by first considering the following standard model describing the evolution of state variables $v$:
\begin{align}
        0 &= \mathcal{R}_a(v), \label{eq:governing_eq}\\ 
        d &= \mathcal{O}(v) + \epsilon_m, \quad \epsilon_m \sim \pi(\epsilon_m).\label{eq:obs_op}
\end{align}
In \eqref{eq:governing_eq}, the governing equations are expressed in residual form, with $a$
representing a model aspect requiring an assumption; examples include constitutive equations, couplings, or even neglected terms.
The calibration data $d$ is then determined by an observation operator $\mathcal{O}$ acting on the state variables, with measurement noise $\epsilon_m \sim \pi(\epsilon_m)$.

MFU representations are embedded into the governing equations as represented by the following modification to~\eqref{eq:governing_eq}--\eqref{eq:obs_op}:
\begin{align}
    0 &= \mathcal{R}_{\xi_\gamma}(v), \label{eq:mfu_gov}\\ 
    d &= \mathcal{O}(v(\xi_\gamma)) + \epsilon_m, \quad \epsilon_m \sim \pi(\epsilon_m), \label{eq:mfu_obs}
\end{align}
where the MFU representation $\xi_\gamma$ replaces or augments $a$ to represent uncertainty in the assumption.
We note that the aim of directly embedding the representation into the governing equations is to enable extrapolative model predictions beyond observational data.
Although the need to extrapolate is relevant to many real-world engineering problems, existing approaches~\cite{kennedy_bayesian_2001,sargsyan_statistical_2015,sargsyan_embedded_2019} to model-discrepancy or model-form error modeling fall short in enabling extrapolation to regimes or scenarios differing significantly from those considered in calibration.
By directly representing the potential source of misspecification, MFU representations can mitigate misspecification's effects on inference; however, they typically cannot fully eliminate misspecification.

While MFU representations provide a way of characterizing uncertainties arising from a model's form, calibrating such uncertainties has historically been challenging.
Since the aim is to characterize functionally irreducible model-form uncertainties, it is desirable that the MFU representation's uncertainty does not concentrate in the limit of infinite data, when misspecification is present.
Thus, standard and generalized Bayesian paradigms are often ill suited, and existing approaches have instead leveraged hierarchical approaches.
Here, the parameters of the MFU representation are modeled as random variables, whose associated probability distributions are themselves parameterized.
The idea behind such an approach is that even if uncertainty in the hyperparameters concentrates, with proper posing of the distributions, the uncertainty description of the MFU parameters will not collapse.
However, there exist several theoretical and practical challenges associated with this hierarchical approach.
Primarily, the hierarchical approach often induces or exacerbates identifiability in the parameters having a strong influence on predictive uncertainties. 
As a result, highly informative hyperpriors are needed for meaningful inference. Although, in practice, we rarely have access to the information or intuition needed to construct informative hyperpriors. 
See~\cite{portone_theoretical_2025} for a more in-depth discussion of the challenges associated with applying hierarchical inference to MFU calibration.

This work explores prediction-oriented inference as an alternative, potentially more intuitive and informative, way to calibrate such MFU representations.
Because prediction-oriented posteriors do not collapse when model misspecification is present, it provides a natural paradigm for representing  model-form uncertainties. 
Additionally, such frameworks align with leveraging MFU representations to enable extrapolative predictions by ensuring predictive distributions assign high probability to calibration data. 
From a practical perspective, having descriptive, but not overly broad (uninformative) posterior predictives confers confidence when extrapolating such uncertainties.
On the other hand, MFU representations can benefit the prediction-oriented inference problem by mitigating misspecification, potentially enabling tighter uncertainty bounds. 
We investigate the suitability of this novel inference approach to inform MFU representations, as well as the influence MFU representations have on inverse problem solutions, in~\Cref{sec:ade}.


%% file: computational_results.tex
\section{Computational Results}\label{sec:comp_res}

Here we demonstrate the results of applying the prediction-oriented framework presented in~\Cref{sec:pip} to systems with features relevant to engineering applications. 
We first consider a polynomial example in~\Cref{sec:poly}, which admits closed-form expressions for the posterior and its predictive.
Thus, we can examine the impacts of increased misspecification and noise on the resulting solutions.
Additionally, in~\Cref{sec:poly}, we investigate the impact of approximating the predictive density through 1) MC estimators and 2) component-wise density estimation, as such approximations may be necessary for nonlinear physics-based models with high-dimensional data.
In~\Cref{sec:ade}, we then apply this framework in the context of model-form uncertainty representations. 
Here, we leverage PDE models of contaminant transport through a domain, where model misspecification arises due to imprecise knowledge of the dispersion process.
We demonstrate how in comparison to standard Bayesian inference, the combination of the MFU representations and prediction-oriented inference can lead to better calibration of uncertainties.

Our inference algorithms and computational models are implemented in the JAX Python library (\url{https://docs.jax.dev/}) using Optax's (\url{https://optax.readthedocs.io/}) Adam implementation with learning rate $10^{-2}$ to minimize the objective function. 
Unless otherwise stated, we take a multivariate normal distribution as our variational family for VI and PVI, with an initial guess of $\mathcal{N}(0,I)$, and use $100$ model evaluations to estimate the ``agreement to data'' term in the objective---either the average log-likelihood for VI or the log-predictive for PVI.
The covariance of the variational family was parameterized by the lower-triangular elements of its Cholesky factor, with a softplus applied to diagonal elements to ensure positive-definiteness.
We use $5000$ training steps for the polynomial example and $10000$ for the contaminant transport example; training steps are chosen such that we observed convergence of the loss (i.e. the loss asymptotes).

\input{polynomial_ex}
\subsection{Contaminant transport through heterogeneous porous media}\label{sec:ade}

We now explore the use of the prediction-oriented inference in the context of calibrating MFU representations.
Here, we consider the evolution of a contaminant concentration field, where the aim is to infer uncertain parameters in the model from noisy concentration measurements at a specific time. 
Let us assume that the true evolution of the contaminant concentration $\cobs(x,t)$ in time $t$ and space $x$ is given by the following generalized advection diffusion equation:
\begin{eqnarray}\label{eq:general_ade}
    \frac{\partial}{\partial t}\cobs(x, t) + u \frac{\partial}{\partial x} \cobs(x, t) &=& \nu_p \frac{\partial^2}{\partial x^2} \cobs(x, t) + \linop \cobs(x, t), \quad x \in [0, L_x=4], \; \\
    \cobs(x, 0) &=& \exp\left( -\frac{1}{2}\left( \frac{(x-0.85)^2}{(0.02)^2}\right)\right), \\
    \label{eq:general_ade_BCs}
    \cobs(0, t) &=& \cobs(L_x,t),
\end{eqnarray}
where $u$ is the bulk velocity of groundwater, $\nu_p$ is the pore-scale diffusion coefficient, and $\linop$ is a general linear operator representing dispersion effects. 
Since the model is linear in $\cobs$ with periodic boundary conditions, the solution admits a Fourier series representation as follows
\begin{eqnarray}
    \cobs(x,t) = \hat{c}^{\text{obs}}_{0}(0) + 2 \Re\left[\sum_{k=1}^\infty \hat{c}^{\text{obs}}_{k}(0) \exp\left( t \left[ -\nu_p k_x^2 - u (ik_x) + \lambda_k \right] \right)e^{i \frac{2\pi k}{L_x} x}\right], \quad k_x = \frac{2\pi k}{L_x},
\end{eqnarray}
where $\hat{c}^{\text{obs}}_k(0)$ are the Fourier coefficients of the initial condition, $\lambda_k$ are the eigenvalues of $\linop$, and $\Re[z]$ denotes the real part of a complex variable $z$.
We truncate the series to $k=257$ and solve on a uniform grid with $N_x=512$.
Synthetic observational data is then generated as follows:
\begin{eqnarray}\label{eq:y_obs_cont}
    y^{\text{obs}}(x_i) = \cobs(x_i, t=0.5, u^{dg}=1, \nu_p^{dg}=0.01) + e_i, \quad e_i \sim \mathcal{N}(0, \sigma_{\text{noise}}^2), 
\end{eqnarray}
where $\sigma_{\text{noise}} = 0.005$, and the data-generating parameters are given by $\param^{dg} = [u^{dg}, \hspace{0.1cm} \nu_p^{dg}]$.
Data is collected at $\datadim = 64$ equally spaced points in $[0, \hspace{0.1cm}4]$, and again the vector of observational data is noted as $\data^{\text{obs}} = [y_1^{\text{obs}}, \dots, y_{\datadim}^{\text{obs}}$].
Data-generating eigenvalues are generated from a direct numerical simulation of a detailed advection-diffusion equation through a heterogeneous porous medium as described in~\cite{portone_representing_2019}.

\begin{figure}
    \centering
    \includegraphics[width=0.9\linewidth]{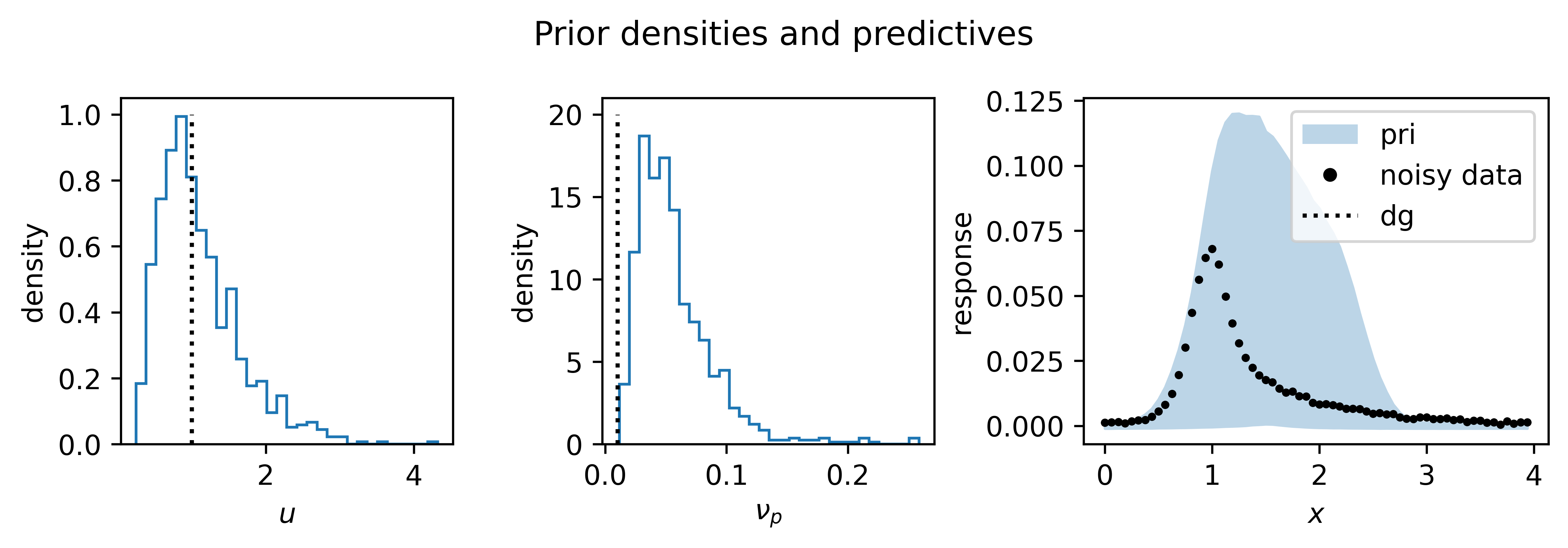}
    \caption{The prior densities of the model parameters $u$ (left) and $\nu_p$ (middle), which are exponentiated from the definitions given in~\eqref{eq:priors_cont} and the corresponding prior predictive $95\%$ credible intervals versus the noisy observational data (right). The dashed lines represent the data-generating (dg) values.}
    \label{fig:ADE_prior_dens}
\end{figure}

To demonstrate how prediction-oriented posteriors can be leveraged alongside MFU representations to improve the predictive power of computational models, let us first assume that the dispersion effects present in the data-generating process are neglected in calibration.
This assumption gives rise to the following inference model defined via the traditional advection-diffusion equation:
\begin{eqnarray}\label{eq:ade_inf}
    \frac{\partial}{\partial t} \cinv(x, t) + u \frac{\partial}{\partial x} \cinv(x, t) &=& \nu_p \frac{\partial^2}{\partial x^2} \cinv(x, t), \quad x \in [0, L_x=4], \\\label{eq:ade_inf1}
    \cinv(x, 0) &=& \exp\left( -\frac{1}{2}\left( \frac{(x-0.85)^2}{(0.02)^2}\right)\right), \\\label{eq:ade_inf2}
    \cinv(0, t) &=& \cinv(L_x,t).
\end{eqnarray}
The aim of the inference problem is to infer the uncertain parameters $\param = [u, \hspace{0.1cm} \nu_p]^{\top}$ from the noisy observational data.
As before,~\eqref{eq:ade_inf}-\eqref{eq:ade_inf2} admits a Fourier series solution given by
\begin{eqnarray}\label{eq:ade_inf_sol}
    \cinv(x, t) = \hat{c}^{\text{inv}}_0(0) + 2\Re \left[\sum_{k=1}^{\infty}  \hat{c}^{\text{inv}}_k(0)\exp\left( t \left[ -\nu_p k_x^2 - u (ik_x)\right] \right)e^{i \frac{2\pi k}{L_x} x}\right], \quad k_x = \frac{2\pi k}{L_x},
\end{eqnarray}
The corresponding statistical model is defined as 
\begin{eqnarray}\label{eq:stat_mod_cont}
    \data = \cinv({\bm x}, t=0.5; \param) + {\bm \epsilon}, \quad {\bm \epsilon} \sim \mathcal{N}({\bm 0}, {\bm \Gamma}_{\text{noise}}),
\end{eqnarray}
where $ {\bm \Gamma}_{\text{noise}} = \sigma_{\text{noise}}^2{\bm I}$, and ${\bm x} = [x_1, \dots, x_{\datadim}]^{\top}$, which gives rise to the likelihood
\begin{eqnarray}\label{eq:lik_cont}
    \likedens(\data | \param) \sim \mathcal{N}\left(\cinv({\bm x}, t=0.5, \param), {\bm \Gamma}_{\text{noise}}\right).
\end{eqnarray}
Since both parameters are positive, we pose the priors via their log-values:
\begin{eqnarray}\label{eq:priors_cont}
    \priordens( \log(u) ) &\sim& \Ncal( \log(1), (0.5)^2), \\\label{eq:priors_cont2}
    \priordens(\log(\nu_p) ) &\sim& \Ncal(\log(0.05), (0.5)^2),
\end{eqnarray}
where standard deviations are selected such that the exponentiated densities for $u$ and $\nu_p$ are physically reasonable; see~\Cref{fig:ADE_prior_dens} for a depiction of the exponentiated priors and corresponding prior predictive $95\%$ credible intervals.

\begin{figure}
    \centering
    \includegraphics[scale=0.7]{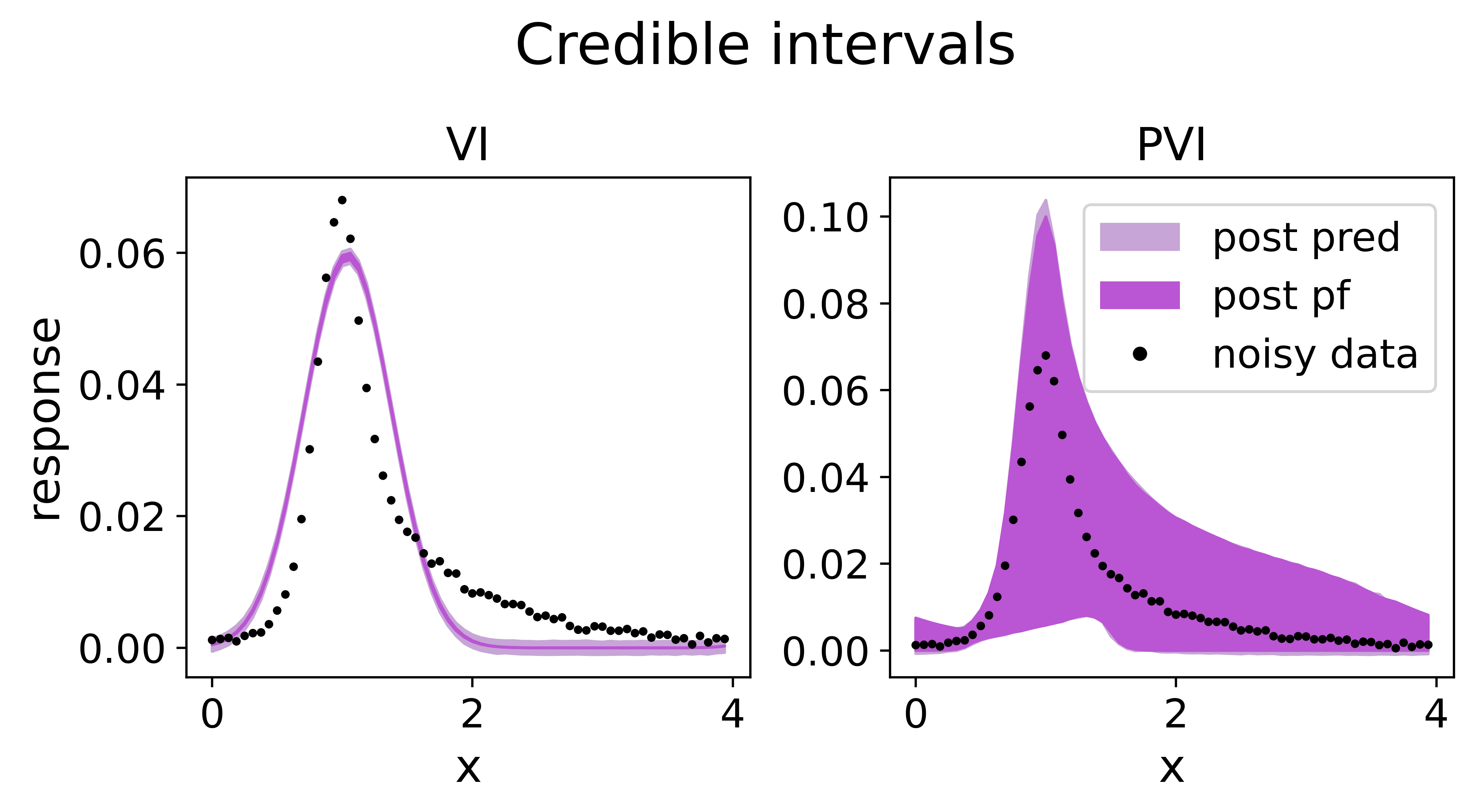}
    \caption{Comparison of the $95\%$ credible intervals of the posterior predictives and pushforwards corresponding to \textbf{standard variational inference (left)} versus \textbf{predictive variational inference (right)} for the inference model defined in~\eqref{eq:ade_inf_sol}.}
    \label{fig:ade_VI_v_PVI}
\end{figure}

\begin{figure}
    \centering
    \includegraphics[width=0.7\linewidth]{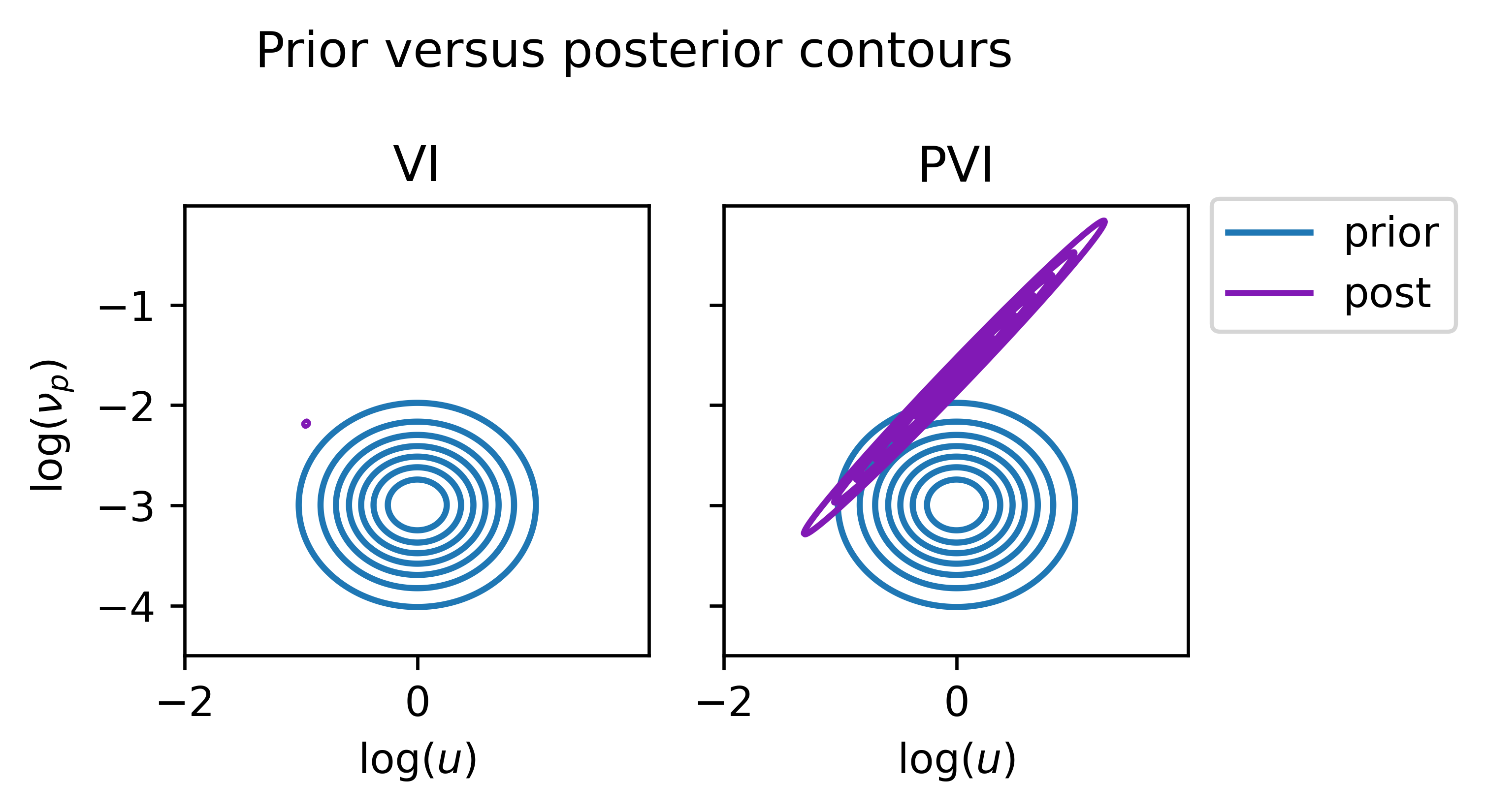}
    \caption{Comparison of prior and posterior contours of the transformed parameters corresponding to \textbf{standard variational inference (left)} versus \textbf{predictive variational inference (right)} for the calibration model defined in~\eqref{eq:ade_inf_sol}.}
    \label{fig:ade_VI_v_PVI_contours}
\end{figure}

\begin{figure}
    \centering
    \includegraphics[scale=0.7]{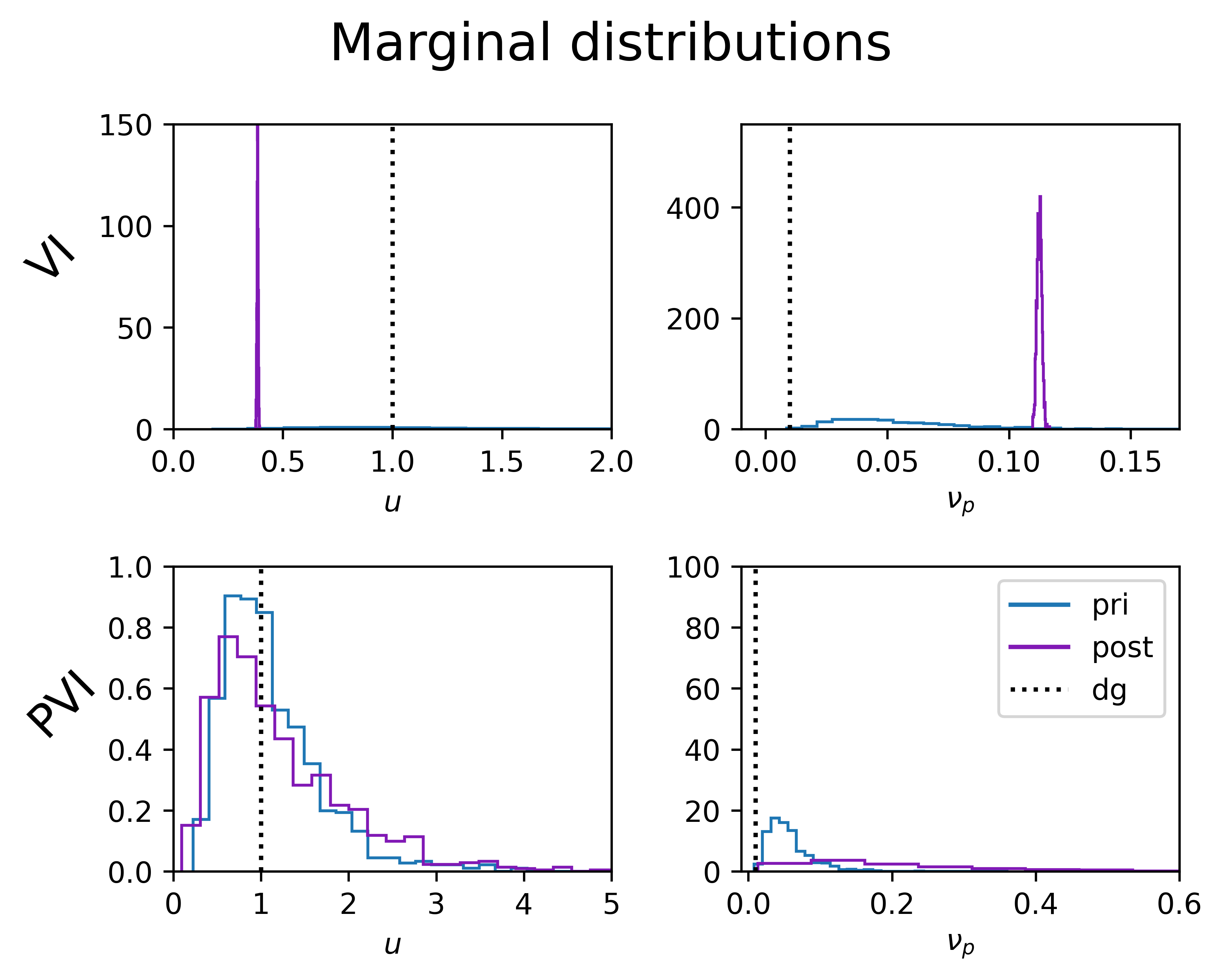}
    \caption{Comparison of prior and posterior marginal distributions versus the data-generating (dg) values for \textbf{standard variational inference (top row)} versus \textbf{predictive variational inference (bottom row)} for the calibration model defined in~\eqref{eq:ade_inf_sol}.}
    \label{fig:ade_VI_v_PVI_marg}
\end{figure}

First, we compare the results of leveraging the inference model defined in~\eqref{eq:ade_inf_sol} in the standard VI versus PVI inverse problems.
Notice from~\eqref{eq:ade_inf_sol} that the parameter-to-observable map is nonlinear in the uncertain parameters. 
Thus, there exists no closed-form representation for the predictive density. 
We therefore leverage a component-wise KDE approximation of the predictive distribution when evaluating the loss in~\eqref{eq:poly_gen_loss}; see~\Cref{alg:kde_pred} and the approximation given in~\eqref{eq:comp_kde} for details.
\Cref{fig:ade_VI_v_PVI} depicts the $95\%$ credible intervals of the posterior predictives and pushforwards corresponding to standard VI (left plot) versus PVI (right plot).
From the left plot of~\Cref{fig:ade_VI_v_PVI}, we see that standard inference underpredicts uncertainty, with the posterior predictives failing to describe the observational data with high probability.
In comparison, the right plot of~\Cref{fig:ade_VI_v_PVI} illustrates how PVI results in a posterior distribution whose pushforward credible intervals encompass the observational data, while still being more informative than prior beliefs. 
In~\Cref{fig:ade_VI_v_PVI_contours}, we directly compare the contours of standard (left plot) and prediction-oriented (right) posteriors relative to the prior for the log-value of the model parameters.
Here, we see that while standard Bayesian inference results in significant concentration of the uncertainty characterization, the prediction-oriented posterior uncertainty characterization did not collapse due to the presence of model misspecification.
In~\Cref{fig:ade_VI_v_PVI_marg}, we further compare the marginals of the standard and prediction-oriented posterior versus the true, data-generating values (horizontal lines).
From~\Cref{fig:ade_VI_v_PVI_marg}, we see that standard VI (top row) results in concentration about biased estimates of the model parameters, i.e. those far from the true data-generating values. 
In comparison, the PVI solution shows significantly less bias in the parameter estimates, although some bias can still be seen in the estimate of $\nu_p$.
Overall, the results indicate that even when significant misspecification is present, PVI can provide better uncertainty characterizations and less biased parameter estimates in comparison to standard VI.

We now investigate the influence of introducing an MFU representation on the solutions to the standard and prediction-oriented inverse problems. 
Here, the MFU representation accounts for dispersion effects, which are known qualitatively to have a nonlocal effect on the evolution of the contaminant; however, the exact nature of this nonlocality is unknown. 
Thus, we pose the following fractional derivative MFU representation of dispersion in the governing advection-diffusion equations:
\begin{eqnarray}\label{eq:frade_inf}
    \frac{\partial}{\partial t} \cmfu(x, t) + u \frac{\partial }{\partial x}\cmfu(x, t) &=& \nu_p \frac{\partial^2 }{\partial x^2}\cmfu(x, t) + \nu \frac{\partial^\alpha }{\partial x^\alpha}\cmfu(x, t) , \quad x \in [0, L_x=4], \\
    \cmfu(x, 0) &=& \exp\left( -\frac{1}{2}\left( \frac{(x-0.85)^2}{(0.02)^2}\right)\right), \\
    \label{eq:frade_inf_BCs}
    \cmfu(0, t) &=& \cmfu(L_x,t).
\end{eqnarray}
Here, the MFU parameters are given by $[\nu, \hspace{0.1cm} \alpha]$, where $\nu$ is a model dispersion coefficient and $\alpha\in(1,2)$ is a fractional power, which for non-integer values yields nonlocal effects on the solution given as 
\begin{eqnarray}\label{eq:frade_sol}
    \cmfu(x, t) = \hat{c}^{\text{mfu}}_0(0) + 2\Re \left[\sum_{k=1}^{\infty} \hat{c}^{\text{mfu}}_k(0) \exp\left( t \left[ -\nu_p k_x^2 - u (ik_x) + \nu(ik_x)^\alpha\right] \right)e^{i \frac{2\pi k}{L_x} x}\right], \quad k_x = \frac{2\pi k}{L_x}.
\end{eqnarray}
Note that in general, the eigenvalues will not directly correspond to the MFU representations, i.e.
$\lambda_k \neq \nu(ik_x)^\alpha$.
So while the MFU representation may mitigate misspecification by representing nonlocal dispersion behavior, it will not entirely eliminate it.

\begin{figure}
    \centering
    \includegraphics[width=0.9\linewidth]{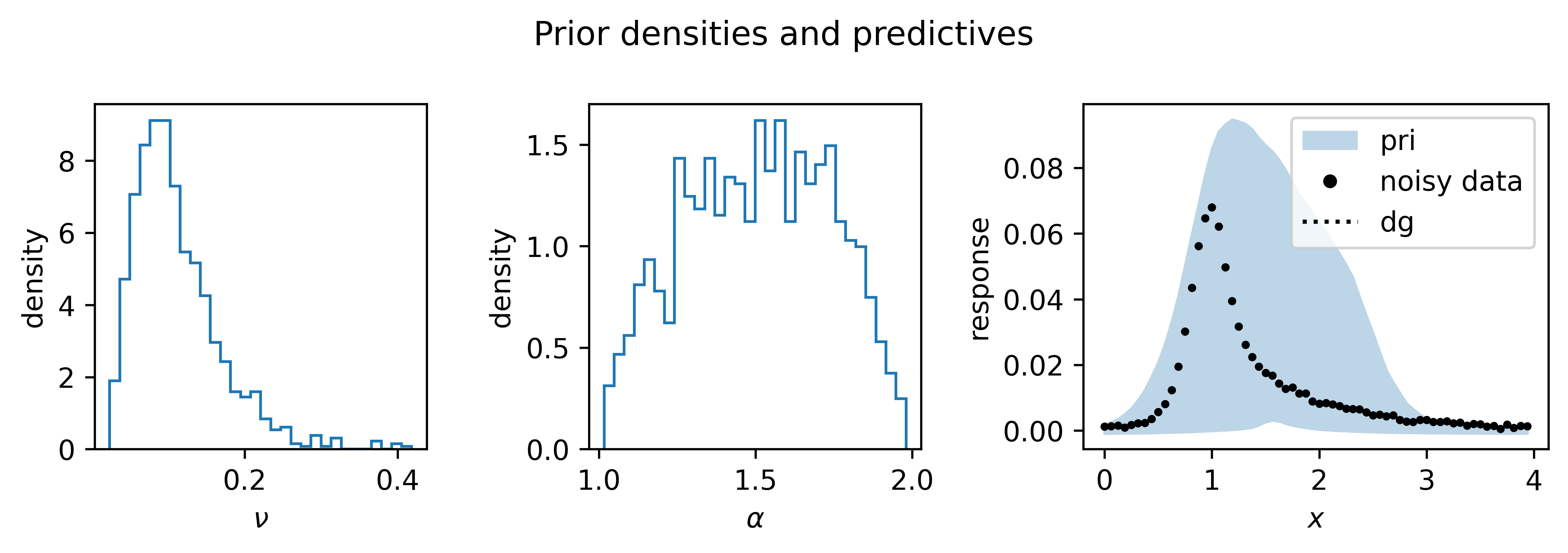}
    \caption{The prior densities of the MFU parameters $\nu$ (left) and $\alpha$ (middle) defined respectively in~\eqref{eq:nu_pri} and~\eqref{eq:alpha_pri}, along with the corresponding prior predictive $95\%$ credible intervals versus the noisy observational data (right).}
    \label{fig:FRADE_prior_dens}
\end{figure}

With the incorporated MFU representation, the aim of the inference problem is to inform 
distributions for model and MFU parameters $\param = [u, \hspace{0.1cm} \nu_p, \hspace{0.1cm} \nu, \hspace{0.1cm} \alpha]$.
We pose priors for $\nu$ and $\alpha$ to respect their bounds.
For $\nu$ we use its log-value:
\begin{eqnarray}\label{eq:nu_pri}
    \priordens(\log(\nu) ) &\sim& \Ncal(\log(0.1), (0.5)^2),
\end{eqnarray}
while for $\alpha$ we pose a prior for the \texttt{probit}\footnote{Here, the \texttt{probit} transformation maps values of $\alpha -1 \in (0, 1)$ to the real line, where a Gaussian prior can be posed.} transform of $\alpha-1$, parameterized here as $g(\alpha)\equiv  \Phi^{-1}(\alpha-1)$, where $\Phi^{-1}(\cdot)$ is the inverse cumulative distribution function of the standard normal, providing
\begin{eqnarray}\label{eq:alpha_pri}
    \priordens\left( g(\alpha)\right) &\sim& \Ncal( g(1.5), (0.8)^2).
\end{eqnarray}
These priors on the transformed values of $\nu$ and $\alpha$ yield densities with nonzero probability in the ranges $(0,.25]$ and $(1,2)$, respectively.
The priors for $u$ and $\nu_p$ are defined as in~\eqref{eq:priors_cont} and~\eqref{eq:priors_cont2}.
Histograms of the prior densities on the MFU parameters, along with the corresponding $95\%$ credible intervals are given in~\Cref{fig:FRADE_prior_dens}.

\begin{figure}
    \centering
    \includegraphics[scale=0.7]{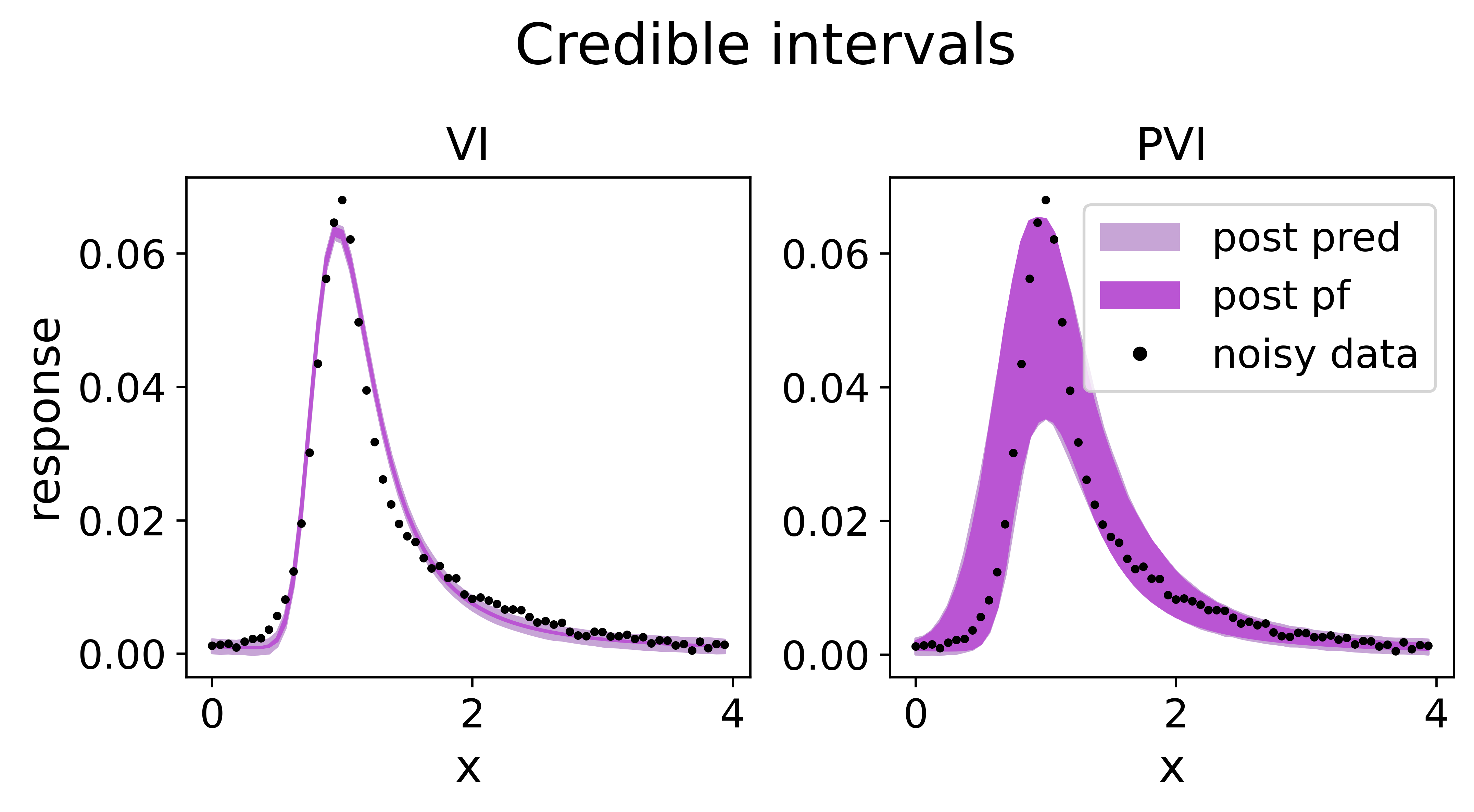}
    \caption{Comparison of the $95\%$ credible intervals of the posterior predictives and pushforwards corresponding to \textbf{standard variational inference (left)} versus \textbf{predictive variational inference (right)} for the inference model defined in~\eqref{eq:frade_sol}.}
    \label{fig:frade_VI_v_PVI}
\end{figure}

To compute the prediction-oriented posteriors, 
we again leverage a component-wise KDE approximation of the predictive distribution when evaluating the loss in~\eqref{eq:poly_gen_loss}.
\Cref{fig:frade_VI_v_PVI} depicts the $95\%$ credible intervals of the posterior predictives and pushforwards corresponding to standard variational inference (left plot) versus predictive variational inference (right plot).
From the left plot of~\Cref{fig:frade_VI_v_PVI}, we see that even with the incorporation of an MFU representation, standard VI underpredicts uncertainty.
Notice, however, in comparing~\Cref{fig:frade_VI_v_PVI} to~\Cref{fig:ade_VI_v_PVI}, we see that although the MFU representation does not fully describe the physics of the data-generating process (defined in~\eqref{eq:general_ade}-\eqref{eq:general_ade_BCs}), its incorporation enables better predictive performance of the model and thus a more informative predictive uncertainty characterization.

\begin{figure}
    \centering
    \includegraphics[scale=0.4]{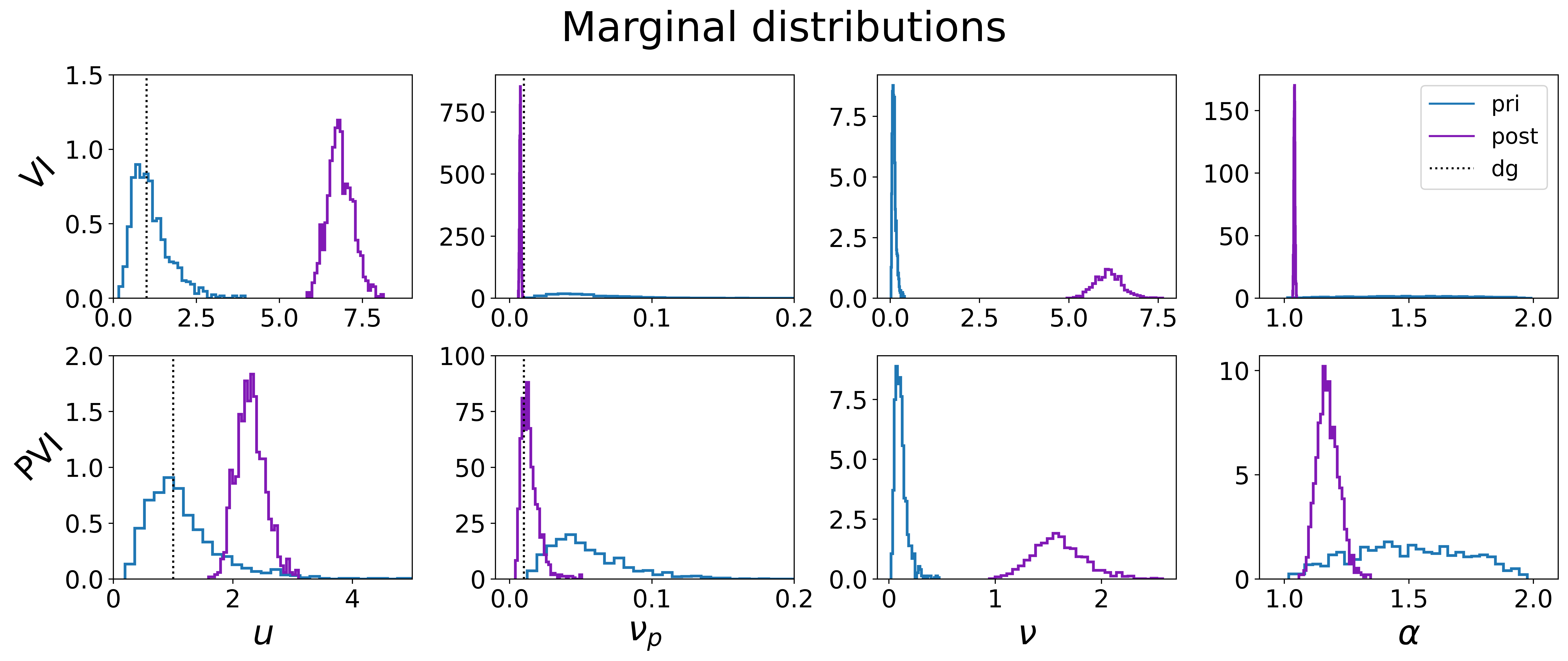}
    \caption{Comparison of prior and posterior marginal distributions versus the data-generating (dg) values for \textbf{standard variational inference (top row)} versus \textbf{predictive variational inference (bottom row)} for the calibration model defined in~\eqref{eq:frade_sol}.}
    \label{fig:frade_VI_v_PVI_marg}
\end{figure}

To understand the impact of incorporation of the MFU representation on the resulting parameter estimates, consider the prior and posterior marginals on the model parameters versus the true, data-generating (dg) values depicted in~\Cref{fig:frade_VI_v_PVI_marg} for standard VI (top row) versus PVI (bottom row).
Note that the MFU parameters do not have data-generating values since they are not present in the data-generating model.
In comparing~\Cref{fig:frade_VI_v_PVI_marg} to~\Cref{fig:ade_VI_v_PVI_marg}, we see that the incorporation of an MFU representation impacts the uncertainty characterizations and bias associated with the model parameters for both VI and PVI. 
Notably, the bias associated with $\nu_p$ is reduced, which intuitively makes sense given that the dispersion (or anomalous diffusion) is, to some degree, accounted for by the MFU term.
In contrast, we see increased bias in the estimates of $u$ when comparing to the left plots of~\Cref{fig:ade_VI_v_PVI_marg}, indicating that the incorporation and calibration of MFU terms can impact the estimation of model parameters.
Comparing the top and bottom rows of~\Cref{fig:frade_VI_v_PVI_marg}, we see the PVI can provide less biased estimates of the model parameters, namely $u$, while also providing MFU parameters whose posteriors are better supported by the prior, namely $\nu$.
%
%
%
Furthermore, a comparison of the posteriors for $\nu_p$ and $\alpha$ between the VI and PVI frameworks shows the PVI framework reduces contraction in the posteriors.

\begin{figure}
    \centering
    \includegraphics[scale=0.8]{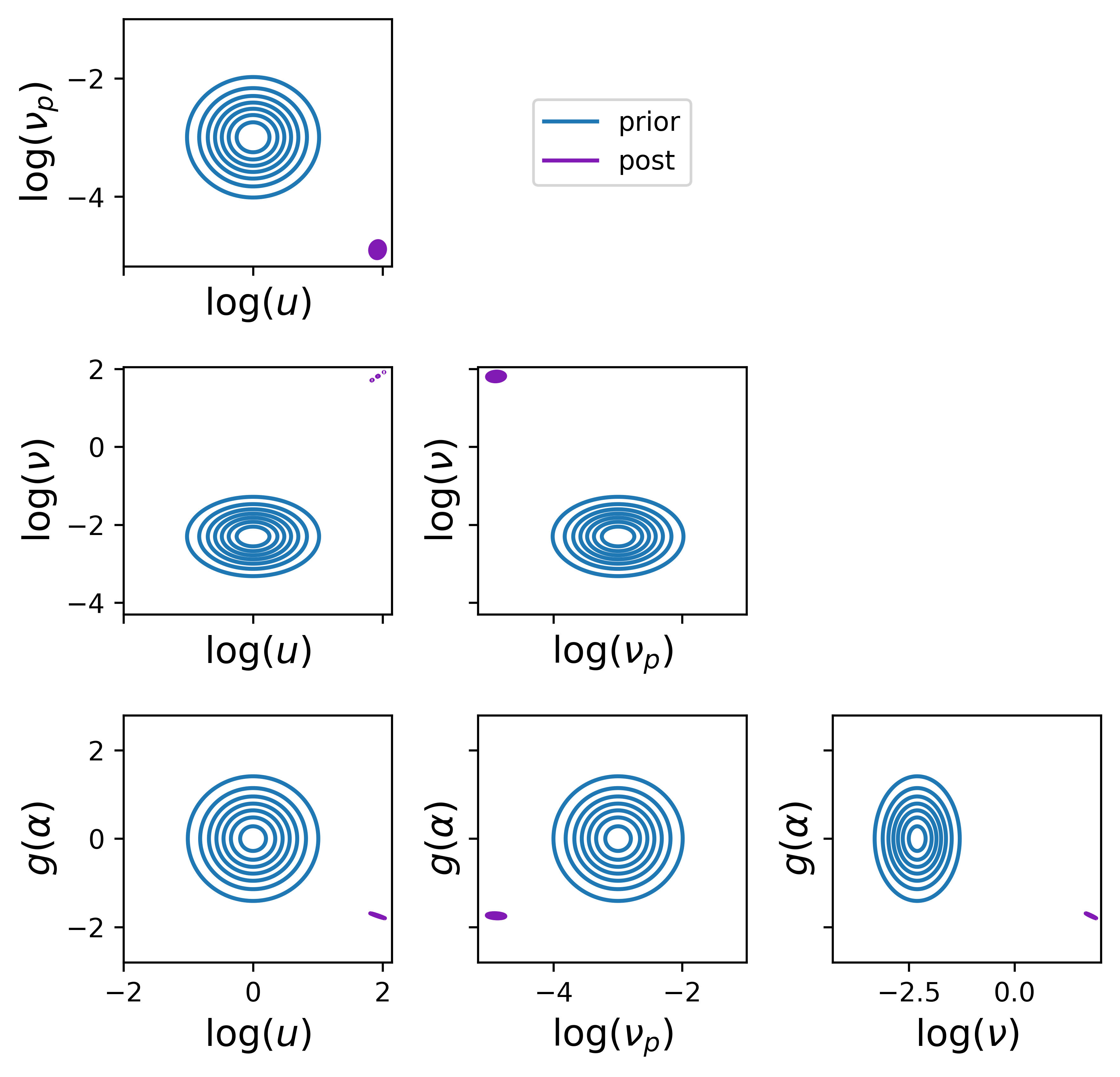}
    \caption{Pairwise comparison of prior and posterior contours of the transformed parameters corresponding to \textbf{standard variational inference} for the calibration model defined in~\eqref{eq:frade_sol}.}
    \label{fig:frade_VI_v_PVI_contours_bayes}
\end{figure}

\begin{figure}
    \centering
    \includegraphics[scale=0.8]{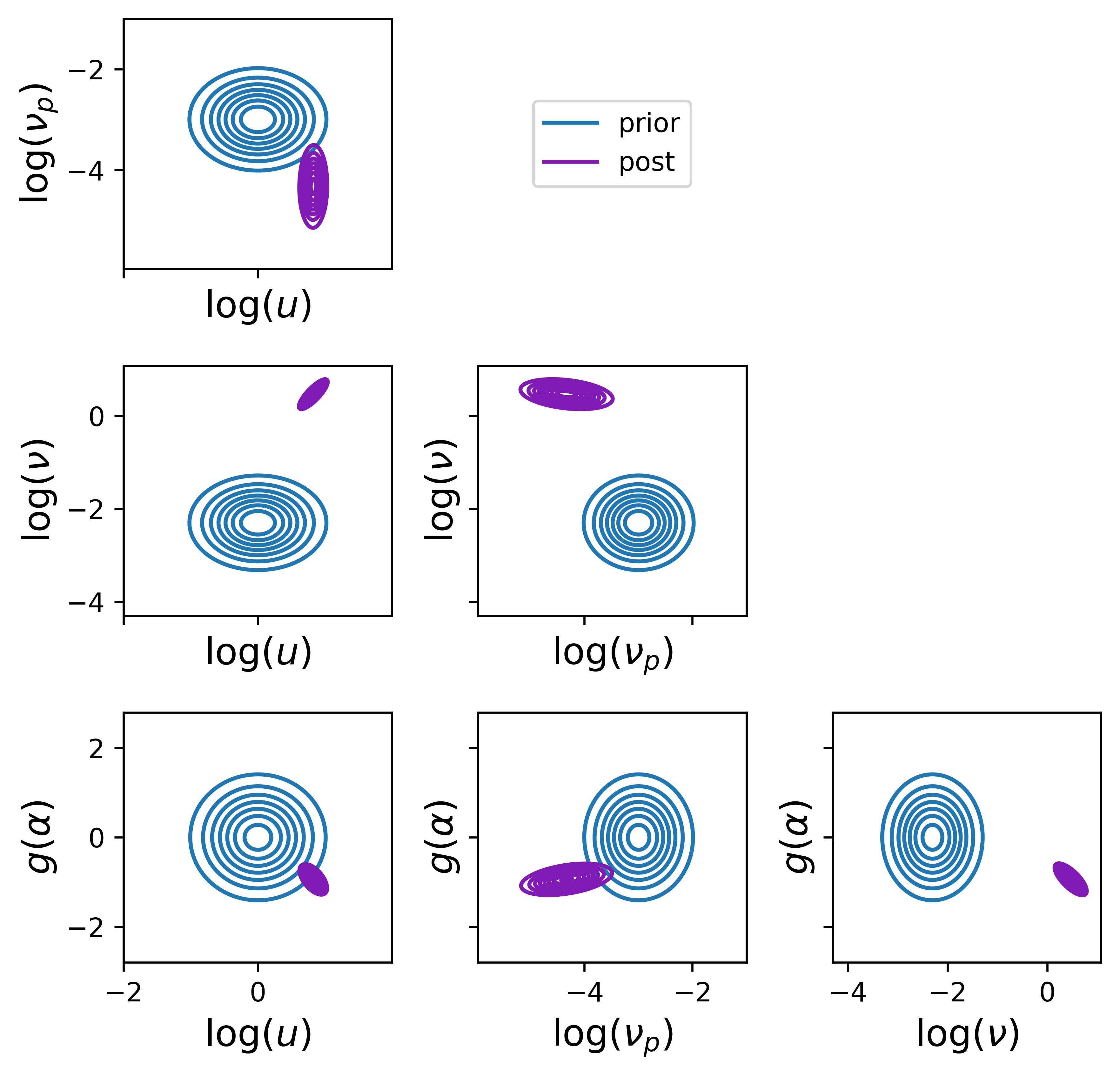}
    \caption{Pairwise comparison of prior and posterior contours of the transformed parameters corresponding to \textbf{predictive variational inference} for the calibration model defined in~\eqref{eq:frade_sol}.}
    \label{fig:frade_VI_v_PVI_contours}
\end{figure}

We further compare pairwise prior and posterior contours of transformed parameters for VI versus PVI respectively in~\Cref{fig:frade_VI_v_PVI_contours_bayes} and~\Cref{fig:frade_VI_v_PVI_contours}.
Here, one can see the reduction in bias and increased uncertainties for the prediction-oriented posteriors in comparison to standard inference.   
Ultimately, while the incorporation of an MFU representation can provide a more predictive model, this alone does not ensure one can meaningfully calibrate uncertainties. 
In fact, the incorporation of the MFU representation can introduce challenges in solving the inverse problem due to the increased dimension of the parameter space and possible spurious dependencies with model parameters that is introduced.
Such challenges, in addition to the persistence of misspecification, can lead to biased parameter densities.
However, the results given in~\Cref{fig:frade_VI_v_PVI,fig:frade_VI_v_PVI_marg,fig:frade_VI_v_PVI_contours_bayes,fig:frade_VI_v_PVI_contours} demonstrate how leveraging prediction-oriented inference frameworks can improve the bias and resulting uncertainty characterizations.

\FloatBarrier


%% file: polynomial_ex.tex
\subsection{Polynomial example}\label{sec:poly}

In this section, we consider polynomial examples that allow us to explore solutions to standard (VI) and predictive variational inference (PVI) problems while leveraging analytical formulas, thus avoiding approximation errors. 
We first present the model and inverse problem formulations, followed by an exploration into the impacts of increasing misspecification and noise on the inverse problem solutions. 
We then investigate the impacts of various approximations of the predictive density, investigating both convergence of MC estimators and component-wise density estimation.

First, let's assume that the true data-generating process is given by the following polynomial model 
\begin{eqnarray}\label{eq:dg_poly}
    y^{\text{obs}}(x_i) = 2x_i^p + 1 + \epsilon_i, \quad \epsilon_i \sim \mathcal{N}(0, \sigma_{\text{noise}}^2),
\end{eqnarray}
where the true data-generating parameters are given by $\param^{dg} = [a^{dg}, \hspace{0.1cm} b^{dg}]^{\top} = [2, \hspace{0.1cm} 1]^{\top}$, the noise in the observational data is given as $\sigma_{\text{noise}}=0.4$, and $p \in (1, 3]$ controls the degree of nonlinearity and hence the degree of misspecification.
The discretization points $x_i$ are computed as $\datadim = 40$ equally spaced points from $[0, \hspace{0.1cm} 2]$.
We denote the vector of observations as ${\bm y}^{\text{obs}} = [y^{\text{obs}}_1, \dots, y^{\text{obs}}_{\datadim}]$.

To understand the impact of misspecification on the solution to the standard and prediction-oriented inverse problems, we assume there is a neglected nonlinearity.
Thus, the assumed inference model is given as 
\begin{eqnarray}\label{eq:lin_mod}
    f(\param; x) = ax + b, \quad x \in [0, \hspace{0.1cm} 2],
\end{eqnarray}
where the uncertain parameters are given by $\param = [a, \hspace{0.1cm} b]^{\top}$. 
The discretized form of the statistical model is then written as
\begin{eqnarray}\label{eq:stat_model_poly}
    {\bm y} = \mathbf{A} \param + {\bm \epsilon}, \quad {\bm \epsilon} \sim \mathcal{N}({\bm 0}, \noise), 
\end{eqnarray}
where $\noise = \sigma^2_{\text{noise}}{\bm I}$ for identity matrix ${\bm I} \in \mathbb{R}^{n_y \times n_y}$, and
\begin{eqnarray}\label{eq:disc_lin_mod}
   \mathbf{A} = \begin{bmatrix}
    x_1 & 1 \\
    x_2 & 1 \\
    \vdots & \vdots \\
    x_n & 1
    \end{bmatrix},
\end{eqnarray}
where ${\bm x} = [x_1,\ldots,x_{\datadim}]^{\top}$ are the discretized locations at which the model is evaluated.
The form of the statistical model in~\eqref{eq:stat_model_poly} gives rise to the likelihood 
\begin{eqnarray}\label{eq:poly_lik}
    \likedens(\data | \param) \sim \mathcal{N}({\bm A}\param, \noise).
\end{eqnarray}

Given the linear inference model in~\eqref{eq:stat_model_poly} and a Gaussian prior $\priordens(\param) \sim \mathcal{N}(\mpri, \Gpri)$, the standard Bayesian posterior is Gaussian $\postdens(\param) \sim \mathcal{N}(\mpost, \Gpost)$, where the closed-form expressions for the mean and covariance are given as 
\begin{eqnarray}\label{eq:mu_post}
    \Gpost &=& \left({\bm A}^{\top}\Ginv{\bm A} + \Gpri^{-1} \right)^{-1},\\\label{eq:G_post}
     \mpost &=& \Gpost \left({\bm A}^{\top}\Ginv{\bm y}^{\text{obs}} + \Gpri^{-1} \mpri \right).
\end{eqnarray}
Notice this solution to the standard Bayesian inverse problem will be precisely the VI solution~\eqref{eq:bayes_opt}, when restricting $\mathcal{P}(\Param)$ to a multivariate Gaussian family.
%

For the PVI problem, we leverage the loss given in~\eqref{eq:poly_gen_loss}, where 
in the linear Gaussian case for $\mathcal{Q} \sim \Ncal(\mub, \Sigb)$, the closed-form expressions for the pushforward and predictive densities are given as
\begin{eqnarray}\label{eq:pf_dens_poly}
    \pi_{\text{pf}}({\bm z}) &\sim& \mathcal{N}\left( \Ab\mub, \Ab\Sigb\Ab^T \right),
    \\\label{eq:pred_dens_poly}
    \pi_{\text{pred}}({\bm z}) &\sim&  \mathcal{N}\left( \Ab\mub, \Ab\Sigb\Ab^T + \noise \right).
\end{eqnarray}
To help differentiate the uncertainty contributions from the model parameters versus the assumed noise model, we evaluate both the posterior predictive and pushforward densities in the results that follow.
Given that the aim is to leverage the model-predicted uncertainties to inform decision making, ensuring meaningful posterior pushforward uncertainty characterizations is a key goal. 
However, when accounting for how well the uncertainty describes the observational data, we leverage the predictive distribution to account for the noise in the observational data.
To visualize these distributions, we plot the component-wise (marginal) $95\%$ credible intervals computed as 
\begin{eqnarray}\label{eq:marg_CI}
    \text{CI}_{95}(x_i) = ({\bm A}{\bm \mu})_i \pm 1.96 \sqrt{\left({\bm \Gamma}_y \right)_{ii}},
\end{eqnarray}
where ${\bm \Gamma}_y$ corresponds to the covariance in~\eqref{eq:pf_dens_poly} or~\eqref{eq:pred_dens_poly} depending on whether the predictive or pushforward is being considered.

\begin{figure}[h]
    \centering
    \includegraphics[width=0.9\textwidth]{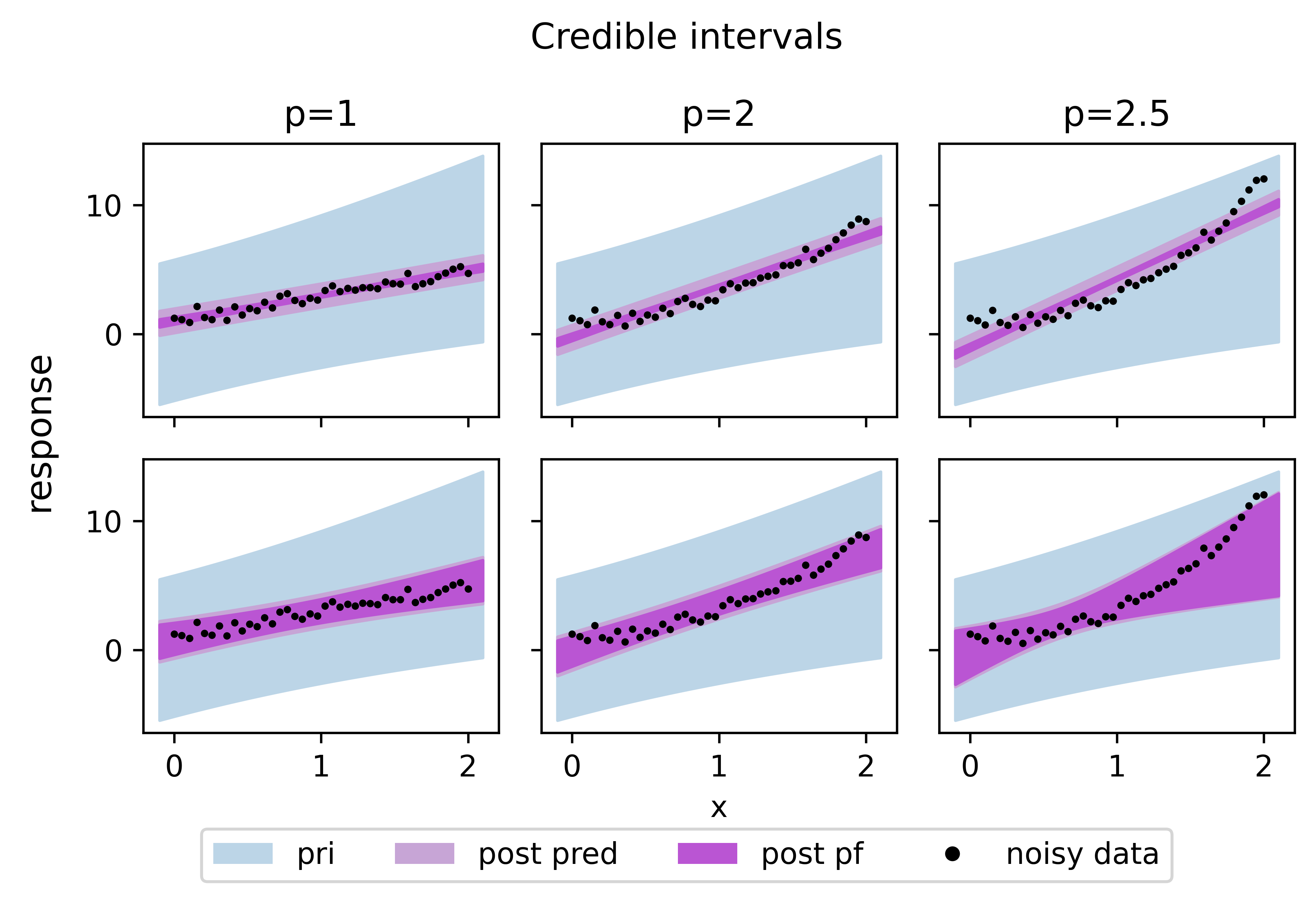}
    \caption{Comparison of the $95\%$ credible intervals of the prior predictive and posterior predictives and pushforwards (pf) corresponding to \textbf{standard variational inference (top row)} versus \textbf{predictive variational inference (bottom row)} for varying levels of misspecification $p$ in~\eqref{eq:stat_mod}.}
    \label{fig:lin_VI_v_GVI_vary_p}
\end{figure}

First, we compare the results of standard VI versus PVI as model misspecification increases. 
Here, the prior is given as 
\begin{eqnarray*}
    \priordens(\param) \sim
    \mathcal{N}\left(\begin{bmatrix}
        3 \\
        0.3 
    \end{bmatrix},  \begin{bmatrix}
        1 & 0 \\
        0 & 2.4 
    \end{bmatrix}\right).
\end{eqnarray*}
In~\Cref{fig:lin_VI_v_GVI_vary_p}, we compare to the $95\%$ credible intervals of the posterior pushforward and predictive densities corresponding to standard VI (top row) versus PVI (bottom row) for increasing misspecification (left to right). 
%
From the top row of~\Cref{fig:lin_VI_v_GVI_vary_p}, we see that standard VI underpredicts uncertainty; even as misspecification increases. That is, as the model becomes harder to fit, predictive uncertainties do not notably increase but rather the mean shifts to minimize the loss given in~\eqref{eq:bayes_opt}.
In comparison, the PVI results given in the bottom row of~\Cref{fig:lin_VI_v_GVI_vary_p} show that by directly penalizing predictive uncertainties, the predictive approach can adapt to the level of misspecification by increasing uncertainties as the inference model becomes harder to fit.
In the limit of infinite data, where one considers the average loss over the realizations of data, PVI will converge to the Bayesian solution only if no misspecification is present~\cite{mclatchie_predictively_2025}.
However, as demonstrated in
the first column of~\Cref{fig:lin_VI_v_GVI_vary_p}, for finite data, the prediction-oriented approach does not return the same solution as the standard approach, but instead has larger posterior and thus predictive uncertainties.

\begin{figure}[h]
    \centering
    \includegraphics[width=0.8\linewidth]{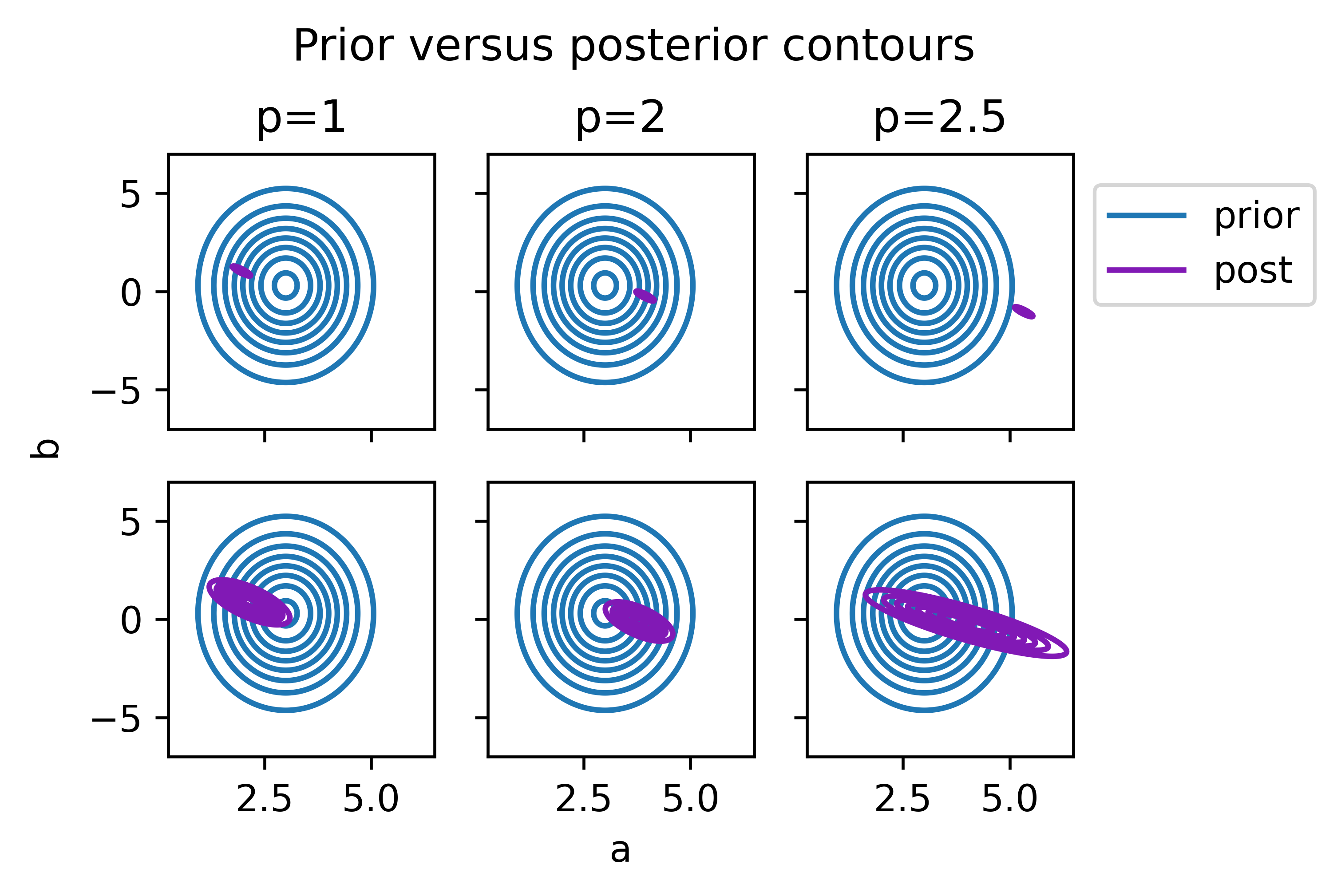}
    \caption{Comparison of prior and posterior contours corresponding to \textbf{standard variational inference  (top row)} versus \textbf{predictive variational inference (bottom row)} for increasing levels of misspecification given by increasing values of $p$ in~\eqref{eq:dg_poly}.}
    \label{fig:post_lin_VI_v_GVI_vary_p}
\end{figure}

Similarly,~\Cref{fig:post_lin_VI_v_GVI_vary_p} depicts the prior and posterior contours corresponding to standard VI (top row) versus PVI (bottom row) for increasing misspecification (left to right).
From the top row of~\Cref{fig:post_lin_VI_v_GVI_vary_p}, we see that as the level of misspecification increases, the posterior mean becomes increasingly biased away from the true data-generating parameters $\param_{dg} = [2, \hspace{0.1cm}1]^{\top}$.
While a similar trend can be seen in the bottom row of~\Cref{fig:post_lin_VI_v_GVI_vary_p}, there is less bias in the posterior mean for the highest level of misspecification ($p=2.5$).
Notably, the impact of using the prediction-oriented framework is an increase in posterior uncertainty that allows the posterior predictive to better describe the observational data.
A key difference to note is that for standard VI, even if misspecification is present, the posterior will asymptotically collapse to a manifold based on the dimensions of the parameter space informed by the parameter-to-observable map. 
From the upper far right plots of~\Cref{fig:post_lin_VI_v_GVI_vary_p,fig:lin_VI_v_GVI_vary_p}, one can see that this manifold may poorly support the true parameter values and lead to poor predictive uncertainties. 
However, for PVI, while the posterior bias increases with misspecification, the predictive uncertainty characterization assigns high probability to the observational data.

\begin{figure}[h]
    \centering
    \includegraphics[width=0.8\linewidth]{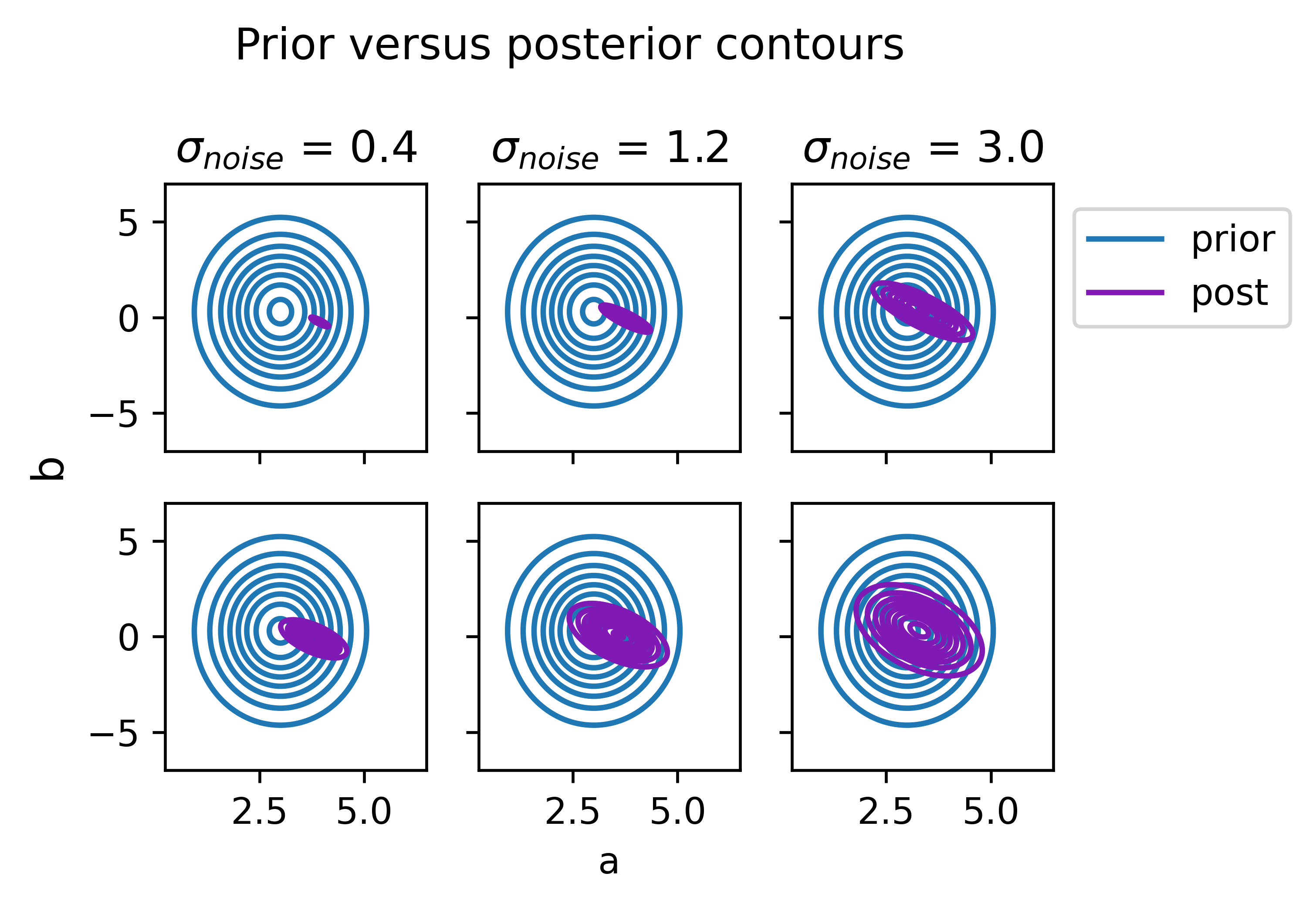}
    \caption{Comparison of prior and posterior contours corresponding to \textbf{standard variational inference (top row)} versus \textbf{predictive variational inference (bottom row)} for a fixed level of misspecification $p=2$ but increasing levels of noise $\sigma_{\text{noise}}$ in~\eqref{eq:stat_mod}.}
    \label{fig:lin_VI_v_GVI_vary_sig_noise}
\end{figure}
\begin{figure}[h]
    \centering
    \includegraphics[width=0.9\linewidth]{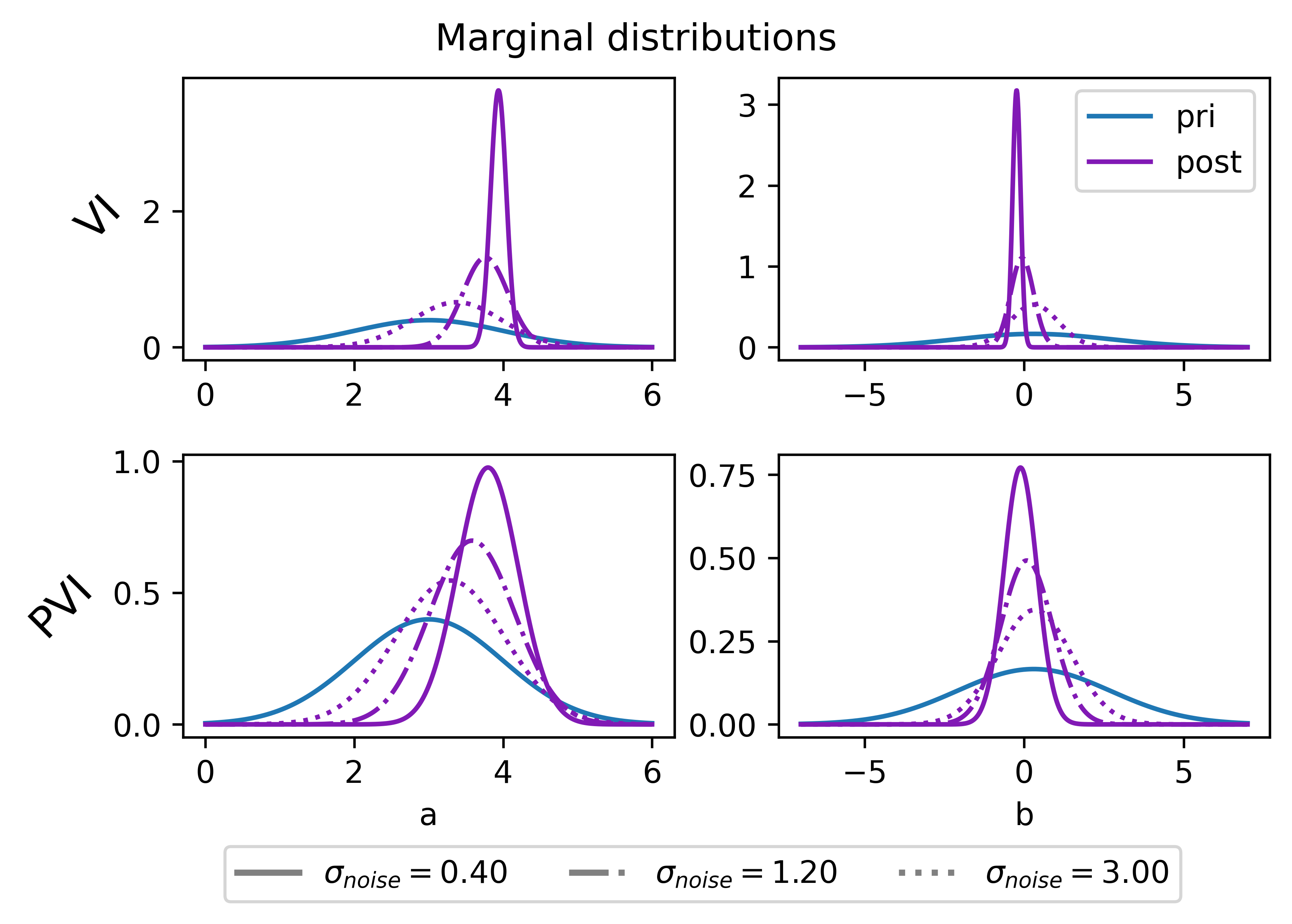} 
    \caption{Comparison of prior and posterior marginals corresponding to 
    \textbf{standard variational inference (top row)} versus \textbf{predictive variational inference (bottom row)} for increasing levels of noise $\sigma_{\text{noise}}$ in~\eqref{eq:stat_mod} (noted in the legend by differing line styles).}
\label{fig:lin_VI_v_GVI_vary_sig_noise_marg_compare}
\end{figure}

Next, we explore the impact of increasing noise levels on the solution to the VI and PVI problems.
In a standard Bayesian paradigm, increasing the noise effectively reduces the informativeness of the data, resulting in a downweighting of the likelihood in~\eqref{eq:bayes_stand} and relative increase in influence of the prior.
Because the PVI problem still adheres to a similar optimization objective---with components representing adherence to observational data as well as prior beliefs---one similarly sees an increase in parameter uncertainties with increased noise as the prior becomes more influential. 
First, consider~\Cref{fig:lin_VI_v_GVI_vary_sig_noise}, which depicts the $95\%$ credible intervals for the posterior predictive and pushforward densities corresponding to standard VI (top row) versus PVI (bottom row) as noise levels increase (left to right). 
Note, that both the observational data (generated according to~\eqref{eq:dg_poly}) as well as the assumed likelihood model in~\eqref{eq:poly_lik} leverage the increased noise. 
Here, we can see that both the VI and PVI, predictive uncertainties increase as the noise increases. 
%
To directly compare the inverse problem solutions as the noise level increases, consider~\Cref{fig:lin_VI_v_GVI_vary_sig_noise_marg_compare,fig:lin_VI_v_GVI_vary_sig_noise}, which respectively depict the marginal densities and contours of the posteriors relative to the prior. 
From these figures, we see increase in posterior uncertainty with increasing noise level, as the posterior becomes increasingly similar to the prior, with PVI having larger uncertainties for each noise level due to the misspecification induced by setting $p=2$ in~\eqref{eq:dg_poly}.
%

\begin{figure}
    \centering
    \includegraphics[width=0.48\linewidth]{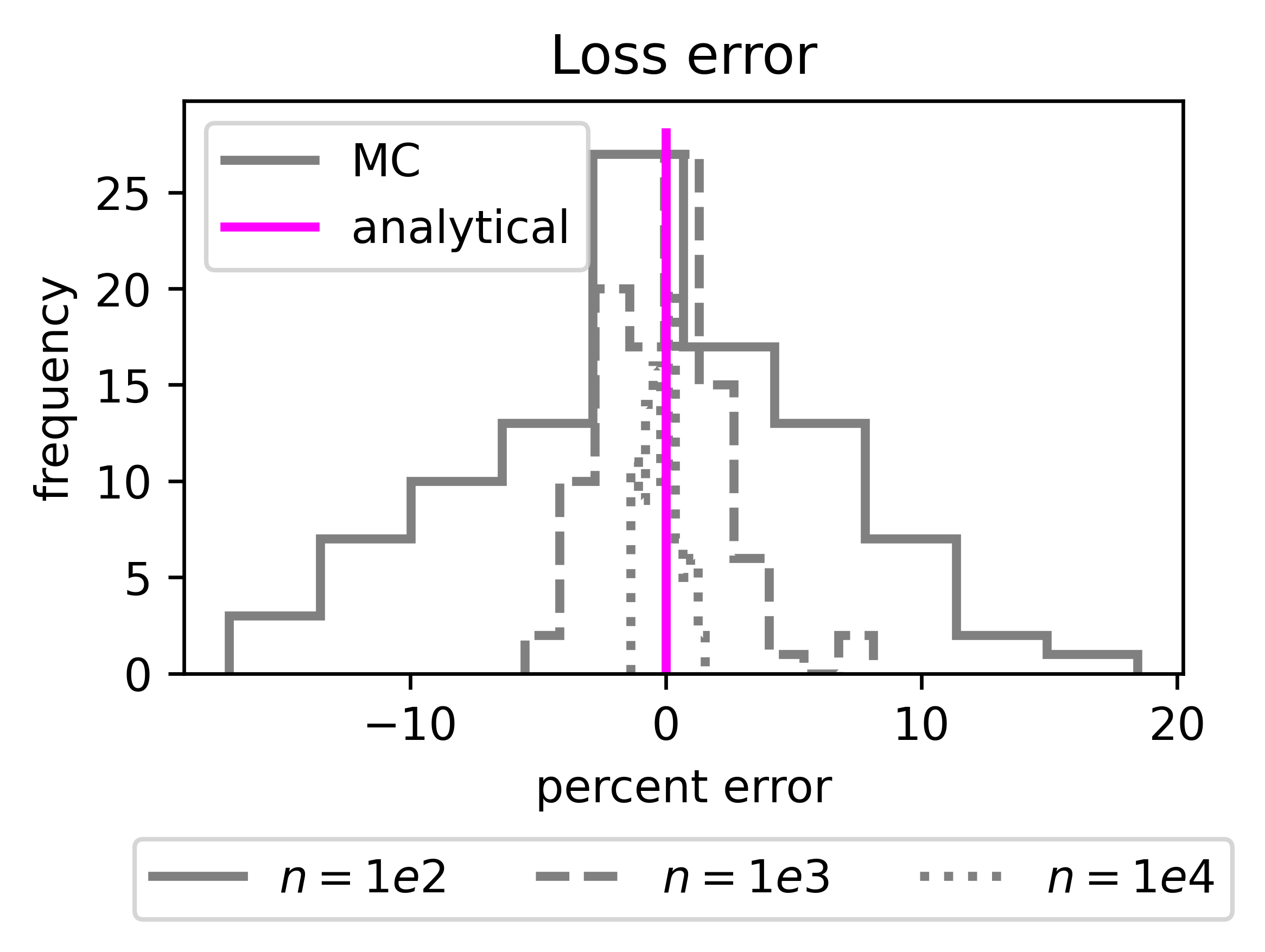}
    \includegraphics[width=0.48\linewidth]{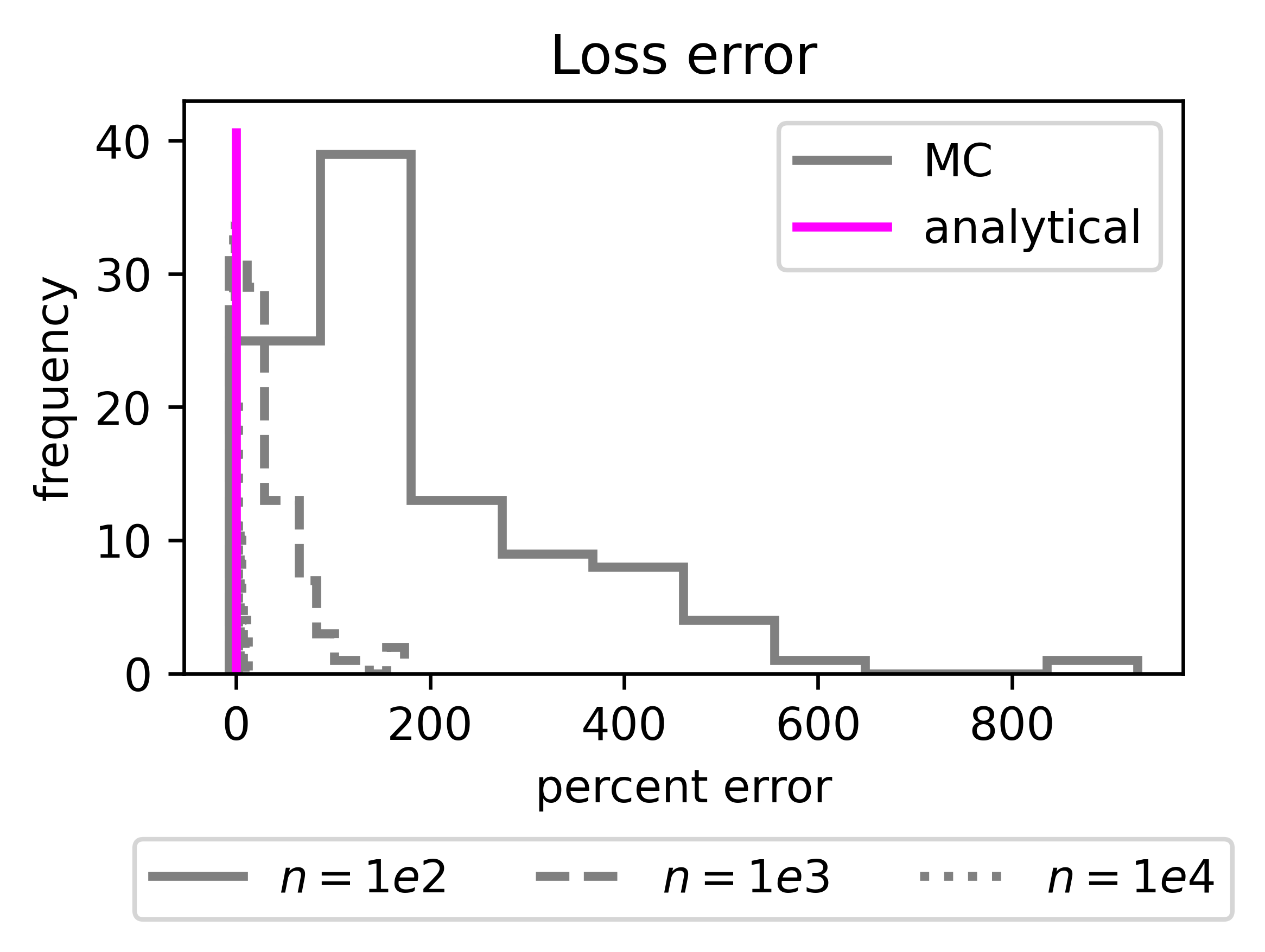}
    \caption{A histogram comparing the percent error in MC-approximated \textbf{variational inference loss (left)}  utilizing~\eqref{eq:exp_lik_MC} (left) versus the \textbf{predictive variational inference loss (right)} utilizing~\eqref{eq:pred_MC} for $100$ replicate MC samples of size $n=10^2, 10^3, 10^4$.
    }
    \label{fig:linear_MC_convergence}
\end{figure}

We now investigate the implication of approximations of the predictive density. 
First, consider that if one has access to a explicit likelihood function, evaluating the standard VI loss in~\eqref{eq:bayes_opt} or the PVI loss given in~\eqref{eq:poly_gen_loss} will require MC approximations of either the expected log-likelihood or the predictive density (given in~\eqref{eq:pred_MC}) at each iteration of the variational optimization problem.
Here, we leverage the closed-form representation of the expected log-likelihood and predictive density for the polynomial example to compare the error associated with MC sampling.
In~\Cref{fig:linear_MC_convergence}, we compare histograms of the percent error in the MC-approximated standard VI loss (left plot) versus the PVI loss (right plot) for $100$ replicate samples of size $n=10^2, 10^3, 10^4$.
The approximated PVI loss is given as the negative log-predictive, where the predictive is approximated according to~\eqref{eq:pred_MC}. 
Similarly, the standard VI loss, given as the negative expected log-likelihood, leverages the following MC estimator 
\begin{eqnarray}\label{eq:exp_lik_MC}
    \mathbb{E}[\log(\likedens(\data | \param))] \approx \frac{1}{n} \sum_{i=1}^{n} \log(\likedens(\data | \param^{(i)})), \quad \param^{(i)} \sim q(\param).
\end{eqnarray}
In both cases, $q \sim \mathcal{N}({\bm 0}, {\bm I})$, where ${\bm I} \in \mathbb{R}^{2\times 2}$ is an identity matrix.
From~\Cref{fig:linear_MC_convergence}, we see that for $n=10^2$, the estimated PVI loss could have as much as $\approx 600\%$ error, whereas the standard VI loss corresponds to at most $\approx 20\%$.
Furthermore, we see that $n=10^4$ MC samples are needed for the PVI loss to converge to the true (analytical) value, which is prohibitive for many problems reliant on expensive forward simulations. 
In general, even when controlling for underflow/overflow numerical issues\footnote{\texttt{logsumexp} provides numerical stability when computing log-marginal likelihoods by utilizing the identity $\log\left(\sum_i e^{w_i}\right) = m + \log\left(\sum_i e^{w_i -m} \right)$, where $m=\max_i w_i$~\cite{jax-logsumexp-doc}.}, approximating the predictive (see~\eqref{eq:pred_MC})
can require significantly more MC samples $n$ than evaluating the marginal log-likelihood.
This instability results from the fact that the likelihood is exponential in the residuals $f(\param) - y$, meaning that the likelihood is nearly $0$ for much of the parameter space, while being very large in a small portion of this space, which is exacerbated for higher-dimensional data.
In comparison, the log-likelihood is quadratic in the residuals, providing faster convergence.
Since the predictive must be estimated at each iteration of the optimization problem, a significantly larger number of likelihood (and hence model) evaluations $n$ may be required to solve the PVI problem in comparison to standard VI, when directly leveraging an explicit likelihood.
%

Next, we explore the impacts of leveraging component-wise density estimation to approximate the predictive from samples, without the need to explicitly evaluate a likelihood function.
While density estimation is an attractive approach as it is likelihood free and can allow for the evaluation of the loss over multiple possible data, density estimation scales poorly with dimension and hence such component-wise approximations may be commonly employed.
Thus, one may need to rely on component-wise estimation of the predictive densities in solving the PVI problem.
For the linear Gaussian problem, we can explore this component-wise assumption analytically, avoiding the need to account for errors induced by density estimation. 
Here, we assume that the predictive density given in~\eqref{eq:pred_dens_poly} is approximated as 
\begin{eqnarray}\label{eq:approx_pred_dens_poly}
    \pi_{\text{pred}}({\bm y}^{\text{obs}}) &\approx& \mathcal{N}\left( \Ab\mub,  \Gpred \right) \\
    &=& \prod_{i=1}^{\datadim}\mathcal{N} \left( (\Ab\mub)_i,  (\Gpred)_{ii} \right). 
\end{eqnarray}
where~$\Gpred = \text{diag}\left(\Ab\Sigb\Ab^T + \Sigb_{\text{noise}}\right)$ ignores the covariances in the predictive density.
This approximation allows us to compute the log-predicted density as a sum of squared error terms as is typical with the standard VI formulation\footnote{Since the noise is independent, the joint Gaussian likelihood factorizes as $\likedens({\bm y}^{\text{obs}} | \param) = \prod_{i=1}^{\datadim} \likedens(y_i^{\text{obs}} | \param)$.
}.

\begin{figure}[h]
    \centering
    \begin{minipage}[c]{0.48\linewidth}
        \centering
        \includegraphics[scale=0.7]{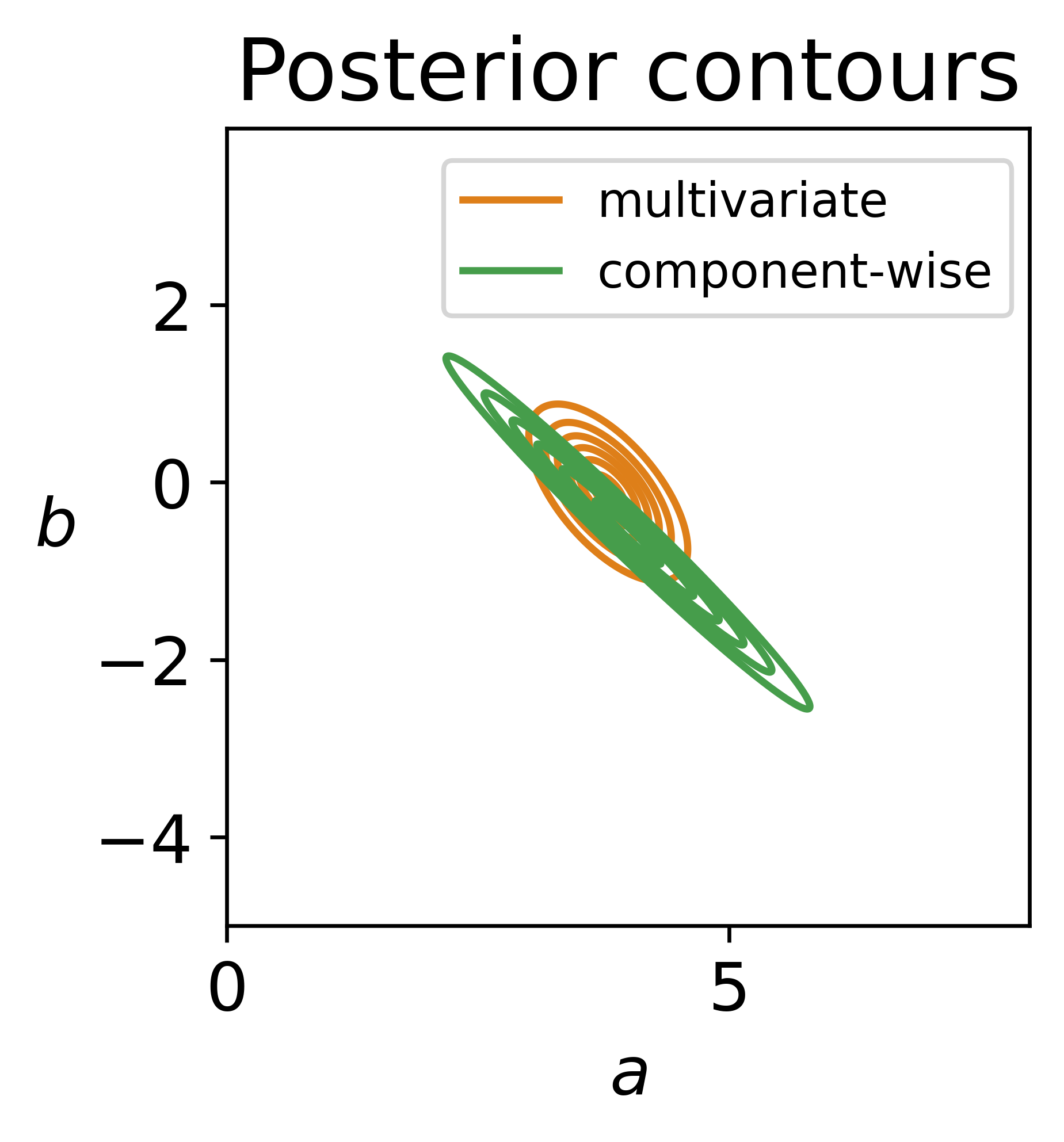}
    \end{minipage}%
    \begin{minipage}[c]{0.6\linewidth}
        \raisebox{0.5\height}{
        \begin{tabular}{lll}
            & Determinant & Trace  \\ \hline
        multivariate  & {\bf 0.02674}    & {\bf 0.435}  \\ \hline
        component-wise & 0.02770    & 1.956   \\
        \hline
        \end{tabular}
        }
    \end{minipage}
    \caption{For \textbf{predictive variational inference}, a comparison of leveraging the full multivariate predictive distribution versus a component-wise approximation in the loss. Contours of the posterior are given in the left plot, with corresponding uncertainties characterized by the trace and determinant of the posterior covariances presented in the table on the right.}
\label{fig_tab:lin_mult_v_component_contours}
\end{figure}

\begin{figure}[h]
    \centering
    \begin{minipage}[c]{0.3\linewidth}
        \centering
        \begin{tabular}{ll}
        \hline
        & Trace  \\ \hline
        multivariate   &    {\bf 15.4}  \\ \hline
        component-wise & 20.3       \\
        \hline
        \end{tabular}
    \end{minipage}%
    \begin{minipage}[c]{0.7\linewidth}
    \includegraphics[width=\linewidth]{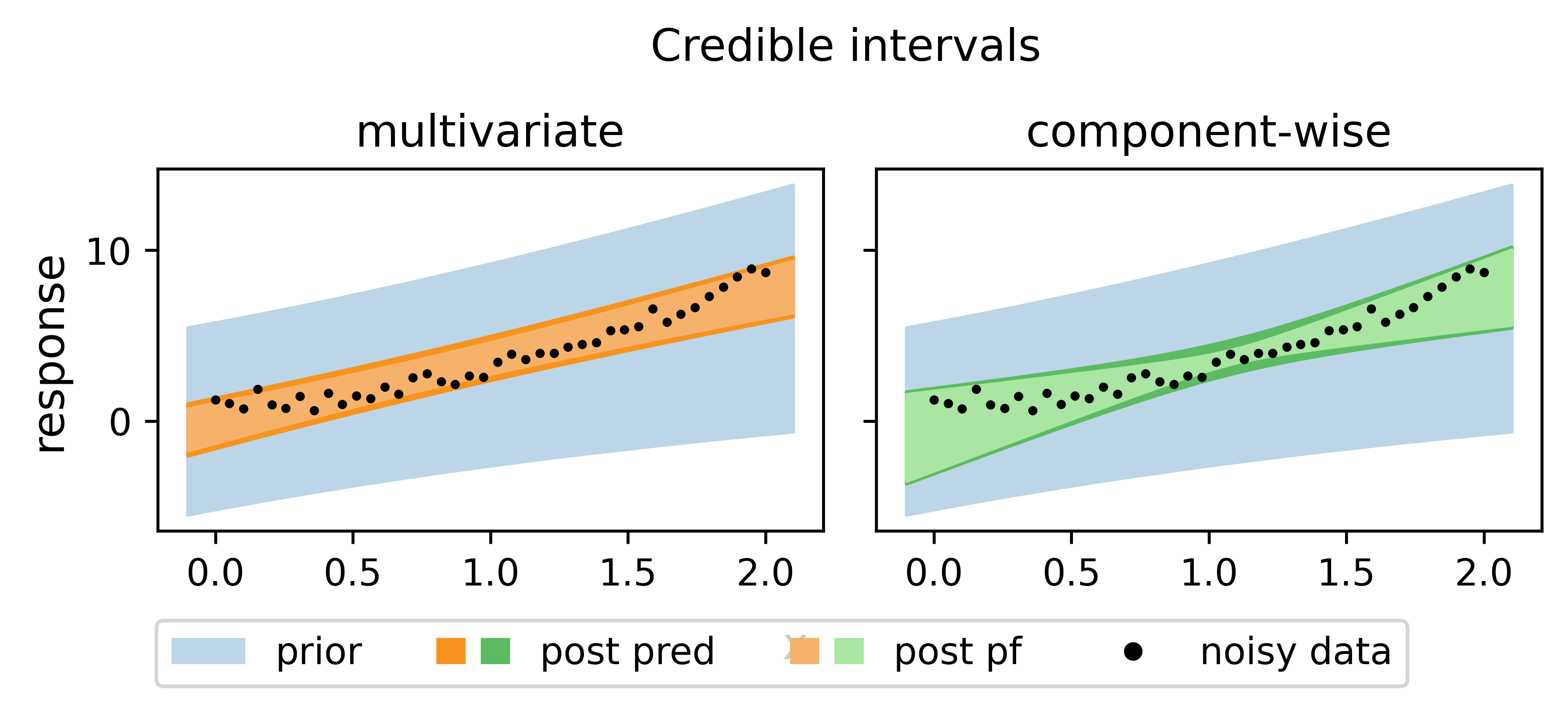}
    \end{minipage}
    \caption{For \textbf{predictive variational inference}, a comparison of the $95\%$ credible intervals of the prior predictive and posterior predictives and pushforwards
    corresponding to a loss formulated with the full \textbf{multivariate Gaussian predictive (left)} versus a \textbf{component-wise approximation (right)}. The corresponding trace of the posterior predictive covariance matrices are presented in the table.}
    \label{fig_tab:lin_mult_v_component_pred}
\end{figure}

For a fixed noise $\sigma_{\text{noise}} = 0.4$ and misspecification level $p=2$,~\Cref{fig_tab:lin_mult_v_component_contours} compares the posterior contours and parameter uncertainties in terms of the trace and determinant of the posterior covariance. 
While the trace provides the sum of parameter variances (ignoring correlations), the determinant is proportional to the hypervolume of the uncertainty ellipsoid. 
From~\Cref{fig_tab:lin_mult_v_component_contours}, we see that parameter uncertainty (characterized by the trace and determinant of the covariance) is reduced when leveraging the full predictive distribution in the loss.
Intuitively, this reduction is because the incorporation of the covariance structure in the predictive better constrains the inverse problem.
The impact on resulting predictive distributions can be seen in~\Cref{fig_tab:lin_mult_v_component_pred}, which compares the predictive uncertainties in terms of the trace of the covariance (left) and predictive credible intervals corresponding to leveraging the full multivariate predictive density (middle) versus the component-wise approximation (right) in the PVI loss given by~\eqref{eq:poly_gen_loss}.
From~\Cref{fig_tab:lin_mult_v_component_pred}, we see that the component-wise assumption results in a posterior predictive with, on average, larger component-wise uncertainties, evidenced by the increased trace of the posterior predictive in the component-wise case\footnote{While one could evaluate the volume of predictive uncertainty via the determinant of the covariance, here the determinant is very small as the data is $40$-dimensional, but the important dimensions impacted by the parameters are $2$-dimensional, making such a comparison less meaningful than the trace comparison.}.
In general, ignoring the covariance structure in the predictive will lead to increased uncertainties as one cannot leverage the underlying correlation structure to better inform the parameters.
Broadly speaking, by only requiring the posterior predictives to describe the data well in a marginal sense, the space of possible solutions is less constrained. 
However, if the aim is to extrapolate uncertainties, ignoring relevant correlations could impact the extrapolative power of the resulting uncertainty characterizations.
%
%
Future efforts could explore the impact of this assumption on extrapolation as well as investigate approaches that enable density estimation in higher dimensions.
\FloatBarrier

%% file: conclusions.tex
\section{Conclusions}\label{sec:conclusions}

This work provides a novel exploration into the use of prediction-oriented Bayesian inference for the calibration of misspecified physics-based models, both with and without the incorporation of model-form uncertainty (MFU) representations.
This optimization-centric inverse problem framework leverages a variational approximation, wherein the prediction-oriented loss is defined as the probability of the calibration data under the log-predictive distribution.
Such a shift from the average log-likelihood-based loss utilized in standard variational inference (VI), i.e. the evidence lower bound (ELBO) loss, results in posteriors whose predictive uncertainties better describe the data, especially in the case where model misspecification is present and for which standard approaches typically underestimate uncertainties.

We apply this predictive variational inference (PVI) approach first to a linear-Gaussian polynomial problem, where closed-form expressions provide the inverse problem solutions and corresponding predictive performances without approximation errors.
Here, we explore the impact of increased misspecification and noise on inverse problem solutions and find that, unlike standard Bayesian inference, the PVI solution adapts to the level of misspecification. 
However, in alignment with standard inference, increased noise decreases the importance of the loss relative to the prior, resulting in prediction-oriented posteriors increasingly conforming to the specified priors as noise levels increase.
Additionally, we explore the impact of 1) Monte Carlo (MC) approximations of the predictive, which require an explicit likelihood and 2) approximations leveraging component-wise density estimation that are amenable to likelihood-free approaches and multivariate data.
For MC approximation, we find that an infeasibly large number of model evaluations may be required per optimization iteration to achieve adequate statistical convergence.
Leveraging component-wise density estimation enables tractable evaluation of the prediction-oriented loss.
However, ignoring the covariance structure of the predictive density alters the constraints of the inverse problem, which can result in larger uncertainties characterized by the trace and determinant of the posterior covariance as well as marginal predictive uncertainties. 

We then apply the PVI framework to partial differential equation models of contaminant transport in heterogeneous porous media, which provide nonlinear parameter-to-observable maps that do not admit closed-form representations of the predictive densities. 
Here, the calibration data is simulated as one realization of high-dimensional spatial data.
We consider two scenarios 1) where model misspecification arises due to contaminant dispersion being neglected and 2) where misspecification is mitigated (but not fully removed) through the incorporation of an MFU representation---comparing the standard VI and PVI solutions for both cases.
For significant model misspecification (scenario 1), we find that the PVI solution provides significantly more reasonable posterior predictive uncertainties in comparison to standard VI, which significantly underpredicts uncertainty.
However, the further incorporation of an MFU representation and subsequent calibration with PVI (scenario 2) results in more concentrated predictive uncertainties while still assigning high probability to the calibration data, highlighting how such an approach can improve a model's predictive performance. 
Furthermore comparing VI and PVI for estimating the model parameters alongside the MFU representation's parameters, we find that PVI reduces bias in the parameter estimates and prevents over-concentration of the posterior. 
Overall, the ultimate aim is to calibrate parameter uncertainties that, when propagated to downstream decision-making quantities of interest, do not provide overconfident predictions while also not being so broad as to be uninformative. 
In this regard, and in terms of describing the observed quantity of interest, we find that PVI provides more meaningful posterior uncertainty characterizations than standard Bayesian inference.

There are several outstanding research directions that could be explored to further the applicability of this approach in the context of large-scale, physics-based modeling.
Future work could explore approaches aimed at improving the computational tractability of applying this prediction-oriented approach, namely more sample efficient and scalable density estimation procedures. 
Methods that admit more expressive density approximations, such as triangular transport or normalizing flows, could enable better posterior approximations.
%
Lastly, since a key aim of the incorporation of MFU representations is improved extrapolation (beyond calibration regimes), studies into how extrapolative goals could be leveraged in the formulation of prediction-oriented losses are of interest.

%% file: acknowledgements.tex
\section*{Acknowledgements}
This work was supported by the Laboratory Directed Research and Development program
(Project 233072) at Sandia National Laboratories, a multimission laboratory managed and
operated by National Technology and Engineering Solutions of Sandia LLC, a wholly owned
subsidiary of Honeywell International Inc. for the U.S. Department of Energy’s National Nuclear
Security Administration under contract DE-NA0003525.
This article has been authored by an employee of National Technology \& Engineering Solutions of Sandia, LLC. The employee owns all right, title and interest in and to the article and is solely responsible for its contents. The United States Government retains and the publisher, by accepting the article for publication, acknowledges that the United States Government retains a non-exclusive, paid-up, irrevocable, world-wide license to publish or reproduce the published form of this article or allow others to do so, for United States Government purposes. The DOE will provide public access to these results of federally sponsored research in accordance with the DOE Public Access Plan \url{https://www.energy.gov/downloads/doe-public-access-plan}.